\documentclass{ws-ijmpa}
\pdfoutput=1
\usepackage{hyperref}
\usepackage{graphicx}
\usepackage{simplewick}
\usepackage{subfigure}
\usepackage[super,compress]{cite}
\usepackage{ulem}
\usepackage{multirow}
\usepackage{siunitx}
\usepackage{slashed}
\usepackage{bm}
\usepackage{ascmac}
\usepackage{amsmath}
\usepackage{amssymb}
\usepackage{here}
\newcommand{\MDWF}{\mathrm{MDWF}}
\newcommand{\OVF}{\mathrm{OVF}}
\newcommand{\PV}{\mathrm{PV}}
\newcommand{\SF}{\mathrm{SF}}
\newcommand{\MSbar}{{\overline{\mathrm{MS}}}}

\def\simge{
    \mathrel{\rlap{\raise 0.511ex
        \hbox{$>$}}{\lower 0.511ex \hbox{$\sim$}}}}
\def\simle{
    \mathrel{\rlap{\raise 0.511ex
        \hbox{$<$}}{\lower 0.511ex \hbox{$\sim$}}}}

\begin{document}
\markboth{Yuko Murakami and Ken-Ichi Ishikawa}
{Construction of Lattice M\"{o}bius Domain Wall Fermions 
in the Schr\"{o}dinger Functional Scheme}

%
\catchline{}{}{}{}{}
%

\title{
CONSTRUCTION OF LATTICE M\"{O}BIUS DOMAIN WALL FERMIONS
IN THE SCHR\"{O}DINGER FUNCTIONAL SCHEME
}

\author{YUKO MURAKAMI$^{1,b}$ and KEN-ICHI ISHIKAWA$^{1,2,a}$}

\address{$^1$Graduate School of Science, Hiroshima University,\\
         Higashi-Hiroshima, Hiroshima 739-8526, Japan\\
$^2$Core of Research for the Energetic Universe, Hiroshima University,\\
Higashi-Hiroshima, Hiroshima 739-8526, Japan\\
$^{a}$ishikawa@theo.phys.sci.hiroshima-u.ac.jp\\
$^{b}$d152338@hiroshima-u.ac.jp}



\maketitle

\begin{history}
\end{history}

\begin{abstract}

In this paper, we construct
the M\"{o}bius domain wall fermions (MDWFs) 
in the Schr\"{o}dinger functional (SF) scheme for the SU(3) gauge theory 
by adding a boundary operator at the temporal boundary of the SF scheme setup.
Using perturbation theory, we investigate the properties of several constructed MDWFs,
including the optimal type domain wall, overlap, truncated domain wall, 
and truncated overlap fermions.
We observe the universality of the spectrum of the effective four-dimensional 
operator at the tree-level, and fermionic contribution to the universal one-loop beta function
is reproduced for MDWFs with a sufficiently large fifth-dimensional extent.

\end{abstract}

\ccode{PACS numbers:11.15.Ha,11.10.Gh,11.30.Rd}


\section{Introduction}
Chiral symmetry has played an important role in quantum field theories and 
in the low-energy physics of Quantum Chromodynamics (QCD). 
Lattice field theory is a non-perturbative, first-principles framework for 
quantum field theories, and lattice gauge theory has been successfully 
applied to describing the low-energy properties of QCD.

Prior to mid-1990s, lattice QCD simulation faced major disadvantages 
in terms of a lack of chiral symmetry on the lattice arising from Nielsen-Ninomiya's 
no-go theorem, which implies the impossibility of constructing lattice fermions 
with chiral symmetry.
Fortunately, the situation has changed following the development of theories for several lattice fermion, including: perfect action\cite{Hasenfratz:1998jp,Hasenfratz:1998ri},
domain wall \cite{Kaplan:1992bt}, and
overlap fermions\cite{Narayanan:1992wx,Narayanan:1993ss,Narayanan:1993sk}.
The notion of symmetry in these lattice fermions can be summarized in terms 
of the Ginsparg-Wilson relation\cite{Ginsparg:1981bj} 
and lattice chiral symmetry\cite{Luscher:1998pqa}.
Overlap fermions exactly hold lattice chiral symmetry, while 
domain wall fermions hold an approximate symmetry with a finite extent in the fifth-dimension.
Because lattice chiral symmetry is not chiral symmetry, the premise
of the no-go theorem is circumvented. 
Lattice chiral symmetry is associated with benefits found in continuum field theories such as no-fine tuning, no-operator 
mixing, \textit{etc.}, and an early review on lattice chiral fermions can be found 
in Refs.~\citen{Creutz:2000bs}-\citen{Chandrasekharan:2004cn}.

Although lattice chiral fermion simulation cost is high, 
large-scale simulations of lattice QCD 
employing lattice chiral fermions, which can be supported by increasing computer performance, 
have been carried out in recent years~\cite{RECENTQCDRESULTS}.
It is necessary to renormalize the lattice operators in a usual manner. 
Non-perturbative renormalization schemes are preferable for reducing systematic errors in applications to involving QCD and Standard model calculations; hence, non-perturbative renormalization of lattice chiral fermions schemes have been sought.

Several non-perturbative renormalization schemes applicable to lattice field theories have been 
developed, including RI-MOM\cite{Martinelli:1993dq,Martinelli:1994ty},
Schr\"{o}dinger functional\cite{Luscher:1992an,Sint:1993un,Luscher:1993gh,Sint:1995rb} (SF),
and Gradient-flow\cite{Luscher:2010iy,Luscher:2011bx,Fodor:2012td,Fritzsch:2013je,Suzuki:2013gza,Luscher:2013cpa}.
The SF scheme has been successfully applied 
for Wilson type fermions to renormalize lattice operators, coupling, and fermion masses, 
together with the $O(a)$-improvements\cite{Luscher:1992an,Sint:1993un,Luscher:1993gh,Sint:1995rb,
Luscher:1996vw,Luscher:1996ug,Luscher:1996jn,Jansen:1998mx,Capitani:1998mq,
Sint:1998iq,Bode:2001jv, DellaMorte:2004bc,Guagnelli:2005zc,DellaMorte:2005kg,Dimopoulos:2007ht}.
The application of the SF scheme to lattice chiral fermions represents a logical next step for extracting their renormalization factors non-perturbatively. 
The SF scheme employs a finite space-time box with a periodic boundary condition 
in the spatial directions and a Dirichlet boundary condition in the temporal direction;
however the temporal boundary condition contradicts 
necessary conditions for overlap and domain wall fermions\cite{Taniguchi:2006qw,Taniguchi:2004gf},
and it has been explicitly noted that the temporal boundary condition must violate chiral symmetry and the Dirac fermion propagator has a nontrivial anti-commutation relation to $\gamma_5$ at the boundary in the continuum theory\cite{Luscher:2006df}.
Therefore to reproduce the same property as in the continuum theory 
it is necessary to introduce an appropriate modification reflecting the boundary condition to the lattice chiral fermions. 
This means that the lattice chiral symmetry becomes nontrivial in the SF boundary condition \cite{Taniguchi:2006qw,Taniguchi:2004gf,Luscher:2006df}.

The SF scheme setup for lattice chiral fermions is formulated in 
Refs.~\citen{Taniguchi:2006qw,Taniguchi:2004gf,Luscher:2006df,Sint:2007zz,Takeda:2007ga,Takeda:2010ai}.
L\"{u}scher investigated chiral symmetry with regarded to the SF boundary condition 
in the continuum theory and pointed out a condition based on universality arguments 
under which the lattice chiral fermions could acquire a proper continuum limit,
where the overlap fermion operator 
was modified through an explicit violation of the lattice chiral symmetry at the SF boundary~\cite{Luscher:2006df}.
The universality of the overlap fermion with the SF boundary condition
was subsequently investigated perturbatively by Takeda~\cite{Takeda:2007ga},
who, in accordance with the universality argument in Ref.~\citen{Luscher:2006df}, 
introduced a boundary term for the domain wall fermion to formulate 
the SF scheme and investigated the property perturbatively\cite{Takeda:2010ai}.

In this paper, we extend the work done by Takeda~\cite{Takeda:2010ai} to 
the M\"{o}bius domain wall fermions (MDWFs)\cite{Brower:2004xi,Brower:2012vk}.
The MDWF is a generalization of the domain wall fermion 
involving the truncated overlap fermion, optimal domain wall fermion, and overlap fermion as limiting cases. 
One motivation for employing the MDWF is 
to reduce simulation cost through the use of a better approximate lattice chiral symmetry. 
After tuning the parameters contained in its action,
the MDWF requires a smaller extent in the fifth dimension 
than the standard domain wall fermion (SDWF) with the same approximate lattice chiral symmetry. 
When the parameters contained in the MDWF action are chosen optimally, 
which corresponds to the optimal domain wall fermion,
lattice chiral symmetry can be realized numerically. 
Thus, the MDWF enables the performance of simulations under a controlled approximate lattice chiral symmetry in a cost-effective manner\cite{Brower:2012vk,Hashimoto:2014gta}.
By developing the SF scheme for the MDWF, a class of lattice chiral fermions 
applicable to the SF scheme can be covered.

To find the boundary operator for the MDWF,
it is necessary to identify the boundary term that properly breaks the chiral symmetry at the temporal 
boundary in the SF scheme.
The boundary operator should hold the discrete space-time symmetries $C, P, T$, and $\Gamma_5$-Hermiticity,
but it must break the lattice chiral symmetry at the temporal boundary only.
In constructing the boundary term, we observe that the parameters of the MDWF, which depend on the fifth-index, 
must have parity symmetry in the fifth dimension in order to hold discrete symmetry.
In this paper, we investigate the validity of the MDWF with a boundary term constructed by observing 
the continuum limit of the effective four-dimensional operator at the tree-level and the one-loop beta 
function perturbatively.
Preliminary results of this work have been reported in Refs.~\citen{Murakami:2017wcz,Murakami:2014qfa}.

The paper is organized as follows.
In section~\ref{sec:SFandCSYM}, we briefly introduce the 
SF formalism and the boundary condition 
and give the property of the Dirac propagator 
under chiral symmetry with the SF boundary condition\cite{Luscher:2006df}.
Before proceeding to the construction of the MDWF with the SF boundary, 
we clarify the property of the MDWF in the periodic boundary condition in section~\ref{sec:PropMDWFwithPBC}.
The continuum limit of the effective four-dimensional operator~\cite{Kikukawa:1999sy,Kikukawa:1999dk} 
is also examined to identify the normalization and the residual mass~\cite{Capitani:2007ez}
at the tree-level.
Section~\ref{sec:PropMDWFwithSFBC} describes 
the construction of the MDWF operator with the temporal boundary term in the SF.
In section~\ref{sec:TreeANL}, 
we numerically verify the continuum limit of the effective four-dimensional operator and the propagator of the MDWF 
with the SF boundary at the tree-level.
In a preliminary work~\cite{Murakami:2014qfa}, 
we did not subtract the residual mass at the tree-level, resulting in non-universal behavior 
at a small extra-dimensional size and discovered that the residual mass must be subtracted to approach 
the  correct continuum limit even at the tree-level. 
We also investigate the tree-level $O(a)$-improvement with the PCAC relation.
In section~\ref{sec:OneLoopBeta}, 
we verify the fermionic portion of the universal one-loop beta function and, 
in section \ref{sec:CONC},
we conclude the paper.

\section{Schr\"{o}dinger Functional Scheme and Chiral Symmetry}
\label{sec:SFandCSYM}

In the Schr\"{o}dinger Functional (SF) scheme~\cite{Luscher:1992an,Sint:1993un,Luscher:1993gh,Sint:1995rb},
 the temporal and spatial extents are
finite with lengths $T$ and $L$, respectively.
To enforce consistency with the time evolution of the quantum system, 
the Dirichlet boundary condition is imposed in the temporal direction.
The temporal coordinate $x_4$ is in $[0,T]$, while the spatial coordinate is $x_j\in[0,L]$.
As we focus on the properties of the fermionic fields, 
we do not explain the details of the gauge field properties in the SF scheme
in this paper.

We begin with the description of the fermionic fields in the continuum theory.
The fermion field $\psi(x)$ in the SF scheme has the following boundary conditions:
\begin{align}
      P_{+}\psi(x)|_{x_4=0}=0,\quad P_{-}\psi(x)|_{x_4=T}=0,
\label{eq:SFBC}
\\
\bar{\psi}(x)P_{-}|_{x_4=0}=0,\quad \bar{\psi}(x)P_{+}|_{x_4=T}=0,
\label{eq:SFBC2}
\end{align}
with $P_{\pm} = (1\pm \gamma_4)/2$.
We employ the generalized periodic boundary condition in the spatial direction,
\begin{align}
\psi(x_0,\bm{x}+L\hat{j})= e^{i\theta_{j}}\psi(x_0,\bm{x}),
\end{align}
where $\hat{j}$ is a unit vector in the $j$-th direction and $\theta_j (j=1,2,3)$ 
is a real parameter.

Although the mass-less Dirac propagator with the temporal boundary conditions given 
in Eqs.~(\ref{eq:SFBC}) and (\ref{eq:SFBC2})
anti-commutes with the $\gamma_5$ operator
\begin{align}
D\gamma_5 + \gamma_5 D = 0,
\end{align}
the mass-less Dirac propagator 
$S(x,y)$ does not anti-commute with $\gamma_5$ as in (\ref{eq:PropACR}):
\begin{align}
\gamma_5 S(x,y) + S(x,y)\gamma_5 &= 
\int_{z_4=0}d\bm{z} S(x,z)\gamma_5 P_{-}S(z,y) +
\int_{z_4=T}d\bm{z} S(x,z)\gamma_5 P_{+}S(z,y).
\label{eq:PropACR}
\end{align}
This relation implies that a naive chiral symmetry does not hold in the SF scheme~\cite{Luscher:2006df}.

On the lattice, the same SF boundary condition 
cannot be imposed directly on the MDWF field, as the MDWF is defined on a five-dimensional lattice.
One required condition for lattice fermion actions in the SF scheme is that the lattice Dirac propagator
should satisfy Eq.~(\ref{eq:PropACR}) in the continuum limit.
Several modifications to the lattice chiral fermion actions to reproduce the SF boundary condition 
are discussed in Refs.~\citen{Taniguchi:2006qw,Taniguchi:2004gf,Luscher:2006df,Takeda:2010ai}.

\section{Properties of M\"{o}bius domain wall fermions in periodic boundary condition}
\label{sec:PropMDWFwithPBC}
Before introducing the SF scheme for MDWFs, 
we briefly explain the MDWF properties without the SF boundary condition. 
For simplicity, the following discussion assumes that the temporal 
and spatial extents of the lattice box are infinite or finite with 
periodic boundary conditions. 

The MDWF operator\cite{Brower:2004xi,Brower:2012vk} 
 $D_{\MDWF}$ is defined by
\begin{align}
D_{\MDWF}&= \left(
    \begin{matrix}
   D_{1}^{(+)} & D_{1}^{(-)}P_{L} &  0 & 0 & 0 & -m_f D_{1}^{(-)} P_{R} \\
   D_{2}^{(-)}P_{R} & D_{2}^{(+)} & D_{2}^{(-)}P_{L} & 0  & 0 & 0 \\
   0  & D_{3}^{(-)}P_{R} & D_{3}^{(+)} & D_{3}^{(-)}P_{L} & 0 & 0 \\
   0 & 0  & D_{4}^{(-)}P_{R} & D_{4}^{(+)} & D_{4}^{(-)}P_{L} & 0 \\
   0 & 0 & 0  & D_{5}^{(-)}P_{R} & D_{5}^{(+)} & D_{5}^{(-)}P_{L} \\
  -m_f D_{6}^{(-)}P_{L} & 0 & 0 & 0  & D_{6}^{(-)}P_{R} & D_{6}^{(+)} 
    \end{matrix}
\right),
\label{eq:MDWF}
\end{align}
\begin{align}
D_{j}^{(+)} &= D_{W} b_j + 1, \quad
D_{j}^{(-)}  = D_{W} c_j - 1, \quad (j=1,2,\cdots,N_5),
\end{align}
where $P_{L/R}= (1\mp \gamma_5)/2$, 
$D_W$ is the four-dimensional Wilson-Dirac operator with a negative mass $(-m_0)$, 
which corresponds to the domain wall height, and $m_f$ is the fermion mass.
The size in the fifth dimension $N_5$ is assumed to be an even number, and we use $N_5=6$ 
to explicitly display the operator in five-dimensional form throughout this paper. 
The coefficients $b_j$ and $c_j$ are tunable parameters of the MDWF 
chosen to optimize the lattice chiral symmetry with a minimum computational cost 
at a finite $N_5$.
In this paper, we generally use $a=1$ as the lattice spacing ``$a$'' 
while writing $a$ explicitly when dealing with the continuum limit.

We employ the MDWF action $S_q$ defined by
\begin{align}
S_q = \bar{\Psi}D^{(N_5)}_{\MDWF}\Psi + \dfrac{1}{2}\Phi^* D^{(N_5)}_{\PV}\Phi,
\label{eq:MDWFPVaction}
\end{align}
where $\Psi$ is the five-dimensional fermion field and
$\Phi$ the Pauli-Villars field.
$D_{\PV}^{(N_5)}$ is defined by $D_{\mathrm{PV}}^{(N_5)} = D_{\MDWF}^{(N_5)}|_{m_f=1}$ 
which eliminates all except the lightest massive mode from the spectrum.
Integrating $\Psi$ and $\Phi$ out in the path-integral with this action,
the fermionic determinant can be expressed~\cite{Kikukawa:1999sy,Kikukawa:1999dk} as
\begin{align}
  \det[D^{(N_5)}_{\MDWF}/D^{(N_5)}_{\PV}] =  \det[D_{\mathrm{eff}}^{(N_5)}],
  \label{eq:quarkdeterminant}
\end{align}
where $D_{\mathrm{eff}}^{(N_5)}$ is the effective four-dimensional operator.
The explicit form of the effective four-dimensional operator $D_{\mathrm{eff}}^{(N_5)}$ in terms of
the MDWF operator (\ref{eq:MDWF}) is given by
\begin{align}
D_{\mathrm{eff}}^{(N_5)} &\equiv
\epsilon^{T}P^{T}
\left(D_{\PV}^{(N_5)}\right)^{-1}
D_{\MDWF}^{(N_5)} P \epsilon.
\label{eq:EffOP}
\end{align}
The permutation and chiral projection matrix $P$ and 
the restriction vector $\epsilon$ are defined by
\begin{align}
P&=
\left(
    \begin{matrix}
P_L & P_R & 0 & 0 & 0 & 0 \\
  0 & P_L & P_R & 0 & 0 & 0 \\
  0 & 0 & P_L & P_R & 0 & 0 \\
  0 & 0 & 0 & P_L & P_R & 0 \\
  0 & 0 & 0 & 0 & P_L & P_R \\
P_R & 0 & 0 & 0 & 0 & P_L  
    \end{matrix}
\right),
\quad   \epsilon = (1,0,0,0,0,0)^{T}.
\end{align}

In the periodic or infinite volume case, we can obtain the explicit form 
for the four-dimensional operator as~\cite{Borici:1999zw,Borici:1999da,Borici:2004pn,Chiu:2002ir,Brower:2004xi,Brower:2012vk}
\begin{align}
D_{\mathrm{eff}}^{(N_5)} &=
\dfrac{1+m_f}{2} + \dfrac{1-m_f}{2} \gamma_5 R_{N_5}({\cal H}_W).
\label{eq:EffOPEXPL}
\end{align}
The matrix function $R_{N_5}(x)$ in Eq.~(\ref{eq:EffOPEXPL})
is the rational approximation to the signum function
\begin{align}
R_{N_5}(x)&= 
\dfrac{\prod_{j=1}^{N_5}(1+w_j x)-\prod_{j=1}^{N_5}(1-w_j x)}
      {\prod_{j=1}^{N_5}(1+w_j x)+\prod_{j=1}^{N_5}(1-w_j x)}.
\label{eq:approxsignfunc}
\end{align}
The matrix ${\cal H}_W$ in the argument is given by
\begin{align}
{\cal H}_W= \gamma_5 D_W ( a_5 D_W + 2)^{-1}.
\label{eq:KernelOP}
\end{align}
The coefficients $(w_j,a_5)$
have the following relation to the MDWF parameters $(b_j, c_j)$:
\begin{align}
    b_j + c_j &= w_j, \quad b_j - c_j = a_5.
\end{align}
This is one of the parameter choice, as the normalization of ${\cal H}_W$ can be absorbed 
into the definition of $w_j$. 
To be specific, we employ the following parametrizations for the Shamir and overlap kernels:
\begin{align} 
  {\cal H}_W &=\gamma_5 D_W (D_W + 2)^{-1}, \quad\mbox{with $b_j+c_j=w_j, b_j-c_j=1$},
\label{eq:SHAMIRKERNEL}
\end{align}
for the Shamir kernel and
\begin{align} 
  {\cal H}_W &=\gamma_5 D_W,                \quad\mbox{with $b_j=c_j=w_j$},
\label{eq:OVERLAPKERNEL}
\end{align}
for the overlap kernel. We investigate these two kernels throughout the paper.

The effective four-dimensional operator in Eq.~(\ref{eq:EffOPEXPL})
reduces to the overlap operator in the limit $N_5\to \infty$ 
provided by the convergence of the approximation to the signum function 
with a given $(b_j,c_j)$. 
We can define an effective four-dimensional fermion theory with Eq.~(\ref{eq:EffOP}) at a finite $N_5$ 
with the following action;
\begin{align}
S_{q \mathrm{eff}}=\bar{\psi}D_{\mathrm{eff}}^{(N_5)}\psi,
\label{eq:truncatedoverlap}
\end{align}
where $\psi$ is the fermion field on the four-dimensional lattice.
We call this theory as the truncated overlap fermion because $R_{N_5}(x)$ in Eq.~(\ref{eq:approxsignfunc}) is
a truncation of the signum function approximation
and Eq.~(\ref{eq:EffOPEXPL}) is an approximation to the overlap fermion operator at a finite $N_5$.

The MDWF operator generalizes the domain wall type fermion operators, 
which include the standard domain wall,
Bori\c{c}i's domain wall~\cite{Borici:1999zw,Borici:1999da,Borici:2004pn}, and optimal Chiu's domain wall~\cite{Chiu:2002ir} 
fermions.
The overlap operator, $D_{\OVF}$, in the limiting case satisfies 
the Ginsparg-Wilson (GW) relation~\cite{Ginsparg:1981bj}
\begin{align}
  \gamma_5 D_{\OVF} + D_{\OVF}\gamma_5  = \dfrac{2}{1+m_f}
\left(
m_f \gamma_5 +D_{\OVF}\gamma_5D_{\OVF}
\right).
\end{align}

At a finite $N_5$, the effective four-dimensional operator 
$D_{\mathrm{eff}}^{(N_5)}$ does not satisfy the GW relation 
owing to the explicit breaking of the lattice chiral symmetry. 
To investigate the chiral property in the SF scheme at a finite $N_5$,
this explicit breaking must be taken into account. 
Taking the continuum limit of Eq.~(\ref{eq:EffOPEXPL}) at the tree-level, 
we obtain
\begin{align}
aD_{\mathrm{eff}}^{(N_5)} &\to
Z 
\left[ i a \slashed{\partial}
+ am_{\mathrm{res}}
\right],
\\
Z&=\dfrac{(1-a m_f)R_{N_5}(\alpha)}{(am_0)(2-(am_0)a_5)},
\label{eq:Normalization}
\\
am_{\mathrm{res}}&= \left[
\dfrac{1+am_f}{1-am_f}\dfrac{1}{R_{N_5}(\alpha)}-1
\right]\dfrac{(am_0)(2-(am_0)a_5)}{2},
\label{eq:ResMass}
\\
\alpha &= \dfrac{(am_0)}{2-(am_0)a_5}.
\end{align}
There is a residual mass $am_{\mathrm{res}}$ in $D_{\mathrm{eff}}^{(N_5)}$ 
even at the tree-level with $m_f=0$~\cite{Capitani:2007ez}. 
However, the residual mass vanishes in $N_5\to \infty$ as $R_{N_5}(\alpha)\to 1$.

The analysis in this section is based on the explicit form of
the effective four-dimensional operator in Eq.~(\ref{eq:EffOPEXPL}). 
In the next section we introduce an MDWF operator for 
the SF scheme by modifying Eq.~(\ref{eq:MDWF}). 
Although the normalization $Z$ of Eq.~(\ref{eq:Normalization}) and 
the residual mass $am_{\mathrm{res}}$ of Eq.~(\ref{eq:ResMass}) 
are obtained in the periodic boundary condition, 
we use these to renormalize the modified MDWF operator with the SF boundary condition at the tree-level.

\section{M\"{o}bius domain wall fermions with the Schr\"{o}dinger functional boundary condition}
\label{sec:PropMDWFwithSFBC}
The SF scheme is defined in a finite four-dimensional box in which the lattice extents
in the spatial and temporal directions are $N_S=L/a$ and $N_T=T/a$, respectively.
The lattice index in the spatial direction $n_k, k=1,2,3$ covers $0, 1, \cdots, N_S-1$,
while the temporal index $n_4$ covers $1, 2, \cdots, N_T-1$. The fields at 
$n_4=0$ and $n_4=N_T$ are fixed using the Dirichlet boundary conditions (\ref{eq:SFBC}) and (\ref{eq:SFBC2}).

According to the universality argument given by
L\"{u}scher~\cite{Luscher:2006df} and the realization of the SF scheme
for the standard domain wall fermion by Takeda~\cite{Takeda:2010ai}, 
we modify the MDWF operator in Eq.~(\ref{eq:MDWF})
by adding a boundary operator $B_{\mathrm{SF}}$ as follows:
\begin{align}
D^{\mathrm{SF}}_{\MDWF}&= D_{\MDWF} + c_{\mathrm{SF}} B_{\mathrm{SF}},
\label{eq:MDWFSF}
\\
B_{\mathrm{SF}}&=
\left(
    \begin{matrix}
   0 & 0 & 0 & 0 & 0 & - D_{1}^{(-)} B \\
   0 & 0 & 0 & 0 & - D_{2}^{(-)} B & 0 \\
   0 & 0 & 0 & - D_{3}^{(-)} B & 0 & 0 \\
   0 & 0 & D_{3}^{(-)} B & 0 & 0 & 0 \\
   0 & D_{2}^{(-)} B & 0 & 0 & 0 & 0 \\
   D_{1}^{(-)} B & 0 & 0 & 0 & 0 & 0 
    \end{matrix}
\right).
\label{eq:BSF}
\end{align}
The boundary operator $B$ is defined by
\begin{align}
    B(n,m) = \delta_{n,m}\gamma_5(\delta_{n_4,1}P_{-}+\delta_{n_4,N_T-1}P_{+}).
\end{align}
The temporal hopping terms from $n_4=0$ and $n_4=N_T$ in the Wilson-Dirac operator 
contained in Eq.~(\ref{eq:MDWFSF}) are set to zero as the SF boundary condition.
The operators in the action Eq.~(\ref{eq:MDWFPVaction}) are replaced with
Eq.~(\ref{eq:MDWFSF}) in the SF scheme. 
Hereafter we omit the superscript $\SF$ on the operator in the SF scheme to simplify the notation.

$B_{\mathrm{SF}}$ has supports only at $n_4=1$ and $N_T-1$, 
and explicitly violates the domain wall chiral symmetry~\cite{Furman:1994ky} at these points;
we expect, however, that the lattice chiral symmetry is still maintained in the bulk region.
The coefficient $c_{\mathrm{SF}}$ is a nonzero parameter for removing 
the $O(a)$-discretization error from the boundary effect.
Although the form of $B_{\mathrm{SF}}$ is a naive extension of Takeda's 
realization, we require its parity symmetry in the fifth-direction
to maintain the discrete symmetries, $C, P, T$ and 
$\Gamma_5$-Hermiticity.
Although this parity symmetry is not required in the periodic boundary or infinite extent cases,
it seems to be of theoretical benefit in the analysis of the corresponding operator and 
action~\cite{Brower:2004xi,Chen:2014hyy,Ogawa:2009ex,Boyle:2015vda,Furman:1994ky}.

The effective four-dimensional operator for Eq.~(\ref{eq:MDWFSF})
is defined by Eq.~(\ref{eq:EffOP});
however, we could not extract a simple short form for the effective 
four-dimensional operator similar to Eq.~(\ref{eq:EffOPEXPL}). 
As seen from Eq.~(\ref{eq:EffOPEXPL}), the ordering of $b_j$ and $c_j$ are irrelevant 
in the periodic or infinite volume cases 
but affect the form of the effective four-dimensional operator 
in the SF scheme, \textit{i.e.}, a different choice for the ordering yields 
a different effective operator. 
This ordering dependence must disappear in the continuum limit, as the SF scheme is
regularization independent.

At a finite $N_5$, the degree of freedom required for $b_j$ and $c_j$ to optimize 
the lattice chiral symmetry with the SF boundary is halved from 
that without the SF boundary as a result of the parity symmetry constraint.
Therefore, we cannot use the optimal choices for 
$b_j$ and $c_j$ as given, \textit{e.g.}, by Chiu~\cite{Chiu:2002ir}.
Choices for the coefficients are given in Refs.~\citen{Chen:2014hyy,Ogawa:2009ex}
and optimal coefficients with the parity constraint are 
formulated in Ref.~\citen{Chiu:2015sea}.
Instead of the optimal coefficients, however,
we employ quasi-optimal coefficients in which 
the half-order ($N_5/2$) coefficients from the Zolotarev 
optimal coefficients~\cite{Chiu:2002eh,Chiu:2002ir} are duplicated to construct the $N_5$ coefficients.
The property of quasi-optimal choice was surveyed in our previous 
study~\cite{Murakami:2014qfa}, in which we checked that the discrepancies between 
quasi-optimal and optimal coefficients~\cite{Chiu:2015sea} were small for $N_5\simge 10$.

Although we cannot extract the rational approximation $R_{N_5}(x)$ 
with the boundary term $B_{\mathrm{SF}}$, we assume that the same rational 
form is valid 
in the bulk region because the boundary effect becomes small as the temporal extent is increased.
For the same reason, we use this form for the normalization $Z$ and the residual 
mass $am_{\mathrm{res}}$ for the SF boundary.
The choice of coefficients $b_j$ and $c_j$ is still important
to maintain the lattice chiral symmetry in the bulk region;
to determine the corresponding values of the coefficients from the approximation range 
for the signum function, we use the spectrum of the kernel 
operator in Eq.~(\ref{eq:KernelOP}) with the Wilson-Dirac fermion operator 
$D_W$ satisfying the SF boundary condition.

\section{Tree-Level Analysis of M\"{o}bius Domain Wall Fermions
with the SF boundary condition}
\label{sec:TreeANL}
In this section, we investigate the properties of the effective four-dimensional operator
built with the MDWF with the SF boundary term in Eq.~(\ref{eq:MDWFSF}) at the tree-level.
The spectrum, temporal structure of the GW relation, and propagator are all 
investigated at the tree-level.
After checking the universality to the continuum limit, 
we will examine the fermionic contribution to the one-loop beta function 
in the next section.
In this paper, 
we employ the Euclidean form\cite{textbook:rothe} of the Dirac representation\cite{textbook:itzyksonzuber} 
for the Dirac gamma matrices.

The classical background field induced by the SF boundary gauge field must 
be incorporated in the analysis with respect to the one-loop beta function. 
The classical background gauge field\cite{Luscher:1993gh} is given by
\begin{align}
  U_k(n) &= 
\exp\left[ 
i \dfrac{1}{N_T}\left(n_4 \phi'_k + (N_T-n_4)\phi_k\right)
\right],\quad
  U_4(n) = 1.
\label{eq:CFIELD}
\end{align}
This is induced by the boundary field
\begin{align}
  \label{eq:BF}
  U_{k}(n)|_{n_4=0}   = e^{i\phi_k}, \quad
  U_{k}(n)|_{n_4=N_T} = e^{i\phi'_k}, 
\end{align}
  for the spatial gauge fields ($k=1,2,3$) at $n_4=0$ and $N_T$~\cite{Luscher:1993gh}.
The standard choices for $\phi_k$ and $\phi'_k$ defining the SF coupling~\cite{Luscher:1993gh} are
\begin{align}
\phi_k  &= \dfrac{1}{N_S}\mathrm{diag}\left(
\eta\omega_1 - \dfrac{\pi}{3}, 
\eta\omega_2,
\eta\omega_3 + \dfrac{\pi}{3}
\right),
\notag
\\
\phi'_k  &= \dfrac{1}{N_S}\mathrm{diag}\left(
-\eta\omega_1 - \pi,
-\eta\omega_3 + \dfrac{\pi}{3},
-\eta\omega_2 +\dfrac{2\pi}{3}
\right),
\label{eq:BoundaryPhase}
\\
\notag
(\omega_1,\omega_2,\omega_3)& =(1,-1/2,-1/2).
\end{align}
This set of choices induces a nonzero chromo-electric field as the background field.
Another choice for the boundary condition is $\phi_k=0$ and $\phi'_k=0$, 
which is useful for analyzing the Dirac propagator analytically in the continuum theory.

The effective four-dimensional operator is renormalized using 
Eqs.~(\ref{eq:Normalization}) and (\ref{eq:ResMass}) as
\begin{align}
 D_q &= Z^{-1} D_{\mathrm{eff}}^{(N_5)}.
\label{eq:renormalizedDq}
\end{align}
In the following, we will investigate the spectrum, the chiral property in the GW relation,
and the propagator of the renormalized $D_q$ with and without the background field.

\subsection{Spectrum of the effective four-dimensional operator}
In this section, we investigate the spectrum of the Hermitian squared effective four-dimensional operator
$D_{q}^{\dag}D_{q}$ in boxes with $N_T=N_S$ with and without the nonzero background field.
We show that the cases $b_j=c_j=w_j$ correspond to the overlap fermion in the $N_5\to \infty$ limit. 
This choice is a nontrivial check to the standard domain wall fermion because $c_j\ne 0$ 
is the new structure for the MDWF.

The quasi-optimal values for the MDWF parameters $(b_j,c_j)$ are determined by
specifying the approximation range and the order $N_5/2$ of the Zolotarev approximation formula.
We use the spectrum of the kernel operator in Eq.~(\ref{eq:KernelOP}) to determine the approximation range. 
We investigate the lowest and highest eigenvalues of the
kernel operator with the SF background field and $\theta=\pi/5$~\cite{Sint:1995ch}, obtaining
\begin{align}
 0.361893\dfrac{1}{N_S} < a|{\cal H}_W| < 1.0,
\label{eq:SHAMIR_KERNELRANGE}
\end{align}
for the Shamir kernel in Eq.~(\ref{eq:SHAMIRKERNEL}) and
\begin{align}
 0.725254\dfrac{1}{N_S} < a|{\cal H}_W| < 7.0,
\label{eq:OVF_KERNELRANGE}
\end{align}
for the overlap kernel in Eq.~(\ref{eq:OVERLAPKERNEL})
from the asymptotic behavior to the continuum limit. 
The lowest eigenvalue approaches zero 
as we take the continuum limit by fixing $L=aN_S$ as a constant;
this means that the approximation to the signum function becomes worse 
with a fixed $(b_j,c_j)$ when the lower spectrum spills over the approximation range 
in taking the continuum limit.

\begin{table}[t]
\tbl{MDWF parameters $b_j (=c_j)$ corresponding to signum function approximation parameters $w_j=b_j$.}
{
  \begin{tabular}{cc} \toprule
  $j$ & $b_j$  \\ \colrule
  $1,8$ & $0.237921110807$ \\ 
  $2,7$ & $1.265028710883$ \\ 
  $3,6$ & $11.29279846601$ \\ 
  $4,5$ & $60.04391219119$ \\ \botrule
  \end{tabular}
  \label{tab:MBPARAM}}
\end{table}

The MDWF parameters $b_j$ and $c_j$ used in the spectrum test, as obtained from the $N_5=4$ optimal Zolotarev approximation coefficient, are listed in Table~\ref{tab:MBPARAM}.
The signum function approximation range and the accuracy 
of the original coefficients $w_j (j=1,\cdots,4)$ 
are $x \in [0.01,7.00]$ and $|1-R_{4}(x)| \simle 0.324$, respectively. 
As seen from Eq.~(\ref{eq:OVF_KERNELRANGE}), this approximation is valid with $N_S < 72$ 
in the cases with the SF background field and $\theta=\pi/5$.
The signum function with the parity-symmetric coefficients
in the fifth dimension, $w_j (j=1,\cdots,8)$, has an accuracy of 
$|1-R_{8}(x)| \simle 0.072$ (Fig.~\ref{fig:signfunc}). 

\newcommand{\figscale}{0.6}
\begin{figure}[t]
  \begin{center}
  \includegraphics[scale=\figscale,clip]{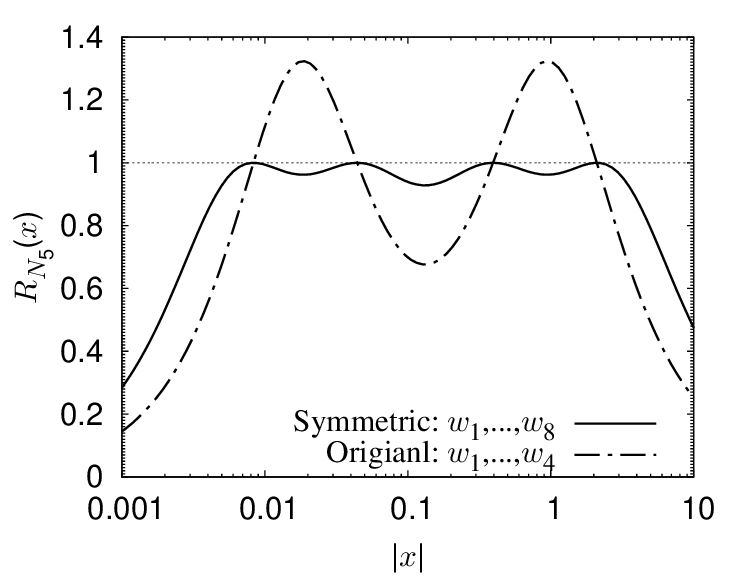}
\caption{Signum function approximation based on Eq.~(\ref{eq:approxsignfunc}).
``Original'' uses $w_1,\cdots, w_4$, which corresponds to the fourth-order optimal 
Zolotarev approximation, 
while ``Symmetric'' uses all coefficients from Table~\ref{tab:MBPARAM}.}
  \label{fig:signfunc}
  \end{center}
\end{figure}

\renewcommand{\figscale}{0.60}
\begin{figure}[t]
 \begin{center}
   \subfigure[Without the background field ($\phi_k=\phi_k'=0$).]{
     \includegraphics[trim=140 10 140 0,scale=\figscale,keepaspectratio,clip]{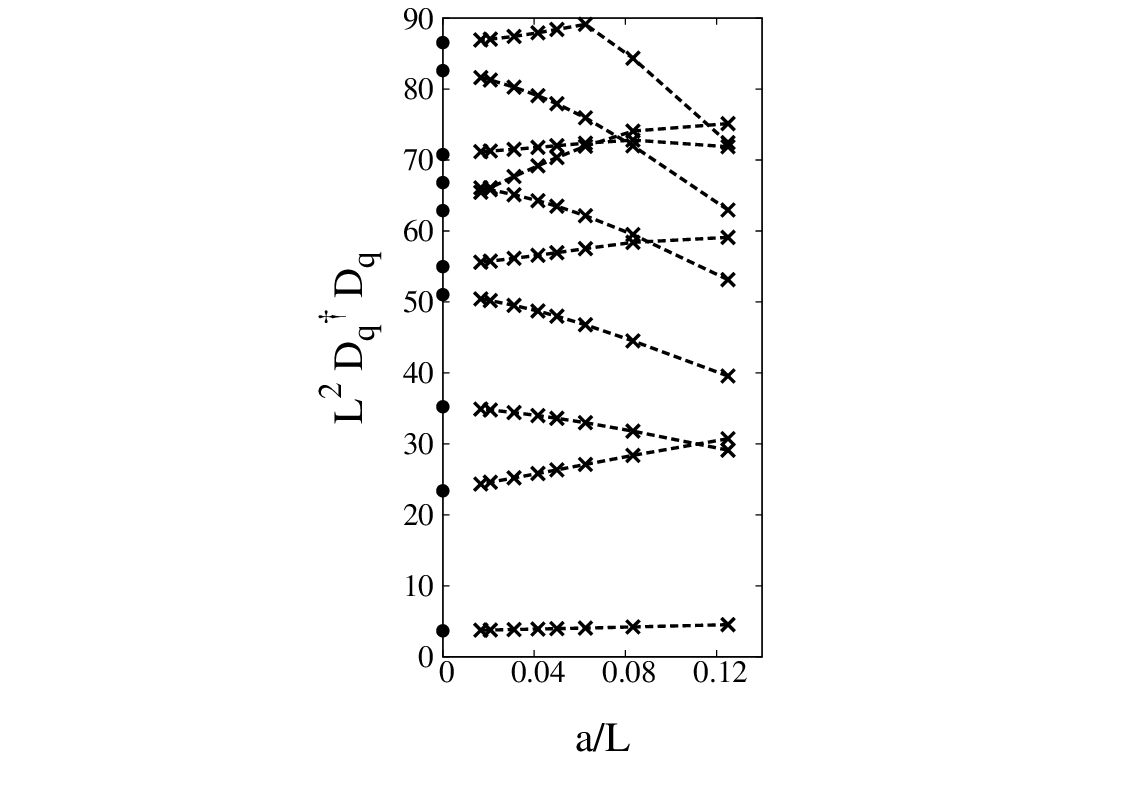}
     \label{fig:EvPcNbgT5}
   }
   \hfill
   \subfigure[With the background field ($\phi_k|_{\eta=0}, \phi_k'|_{\eta=0}$).]{
     \includegraphics[trim= 140 10 140 0,scale=\figscale,keepaspectratio,clip]{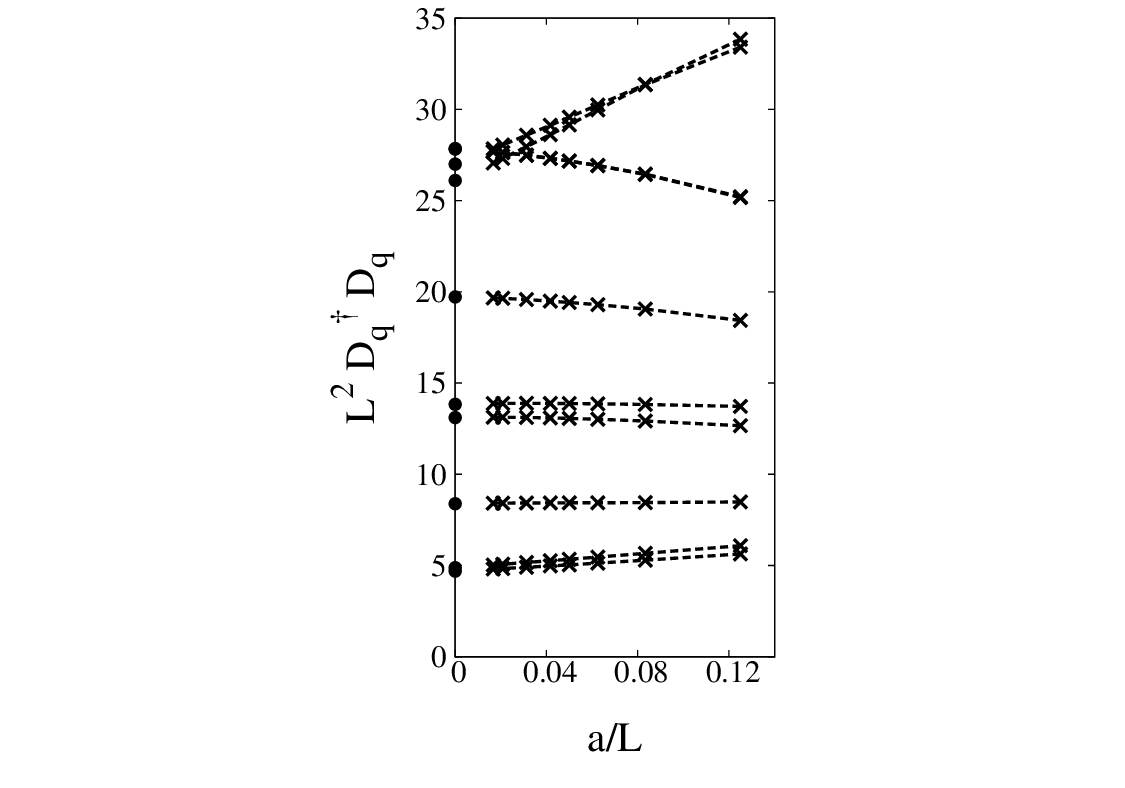}
   \label{fig:EvPcBgT5}
   }
  \end{center}
  \caption{Ten lowest eigenvalues of the MDWF with $c_j=b_j=w_j$, $N_5=8$, $am_{\mathrm{res}}=0$, 
	$\theta=\pi/5$ and $c_{\mathrm{SF}}=1$. 
	The $w_j$ are taken from the optimal Zolotarev approximation with the approximation range $x \in [0.01; 7.00]$ 
	at the given order.}
  \label{fig:EvPcNbg}
\end{figure}

The lowest ten eigenvalues at $am_{\mathrm{res}}=0$ are 
shown in Figure~\ref{fig:EvPcNbg} as a function of $a/L=1/N_S$. 
The boundary coefficient is $c_{\mathrm{SF}}=1$.
The solid circles at $a/L=0$ are the eigenvalues in continuum theory~\cite{Sint:1993un, Sint:1995ch}. 
We use $Z= 0.99659683271$ to normalize Eq.~(\ref{eq:Normalization}) 
and $am_{\mathrm{cr}}=-0.0034031673$, 
where $am_{\mathrm{res}}|_{am_f=am_{\mathrm{cr}}}=0$ from Eq.~(\ref{eq:ResMass}).

The spectrum properly converges to the continuum limit 
even at a smaller $N_5=8$, which supports the use of 
the normalization in Eq. (\ref{eq:Normalization}) and 
residual mass subtraction in Eq. (\ref{eq:ResMass}) even 
with the SF boundary condition.
We confirm the same behavior for other parameters, namely, 
larger values of $N_5$, $am_{\mathrm{res}}\ne 0$, a reverse order of $b_j$, 
and other overlap kernels ($b_j+c_j=1$) and ($b_j=1, c_j=1$).

\subsection{Dirac propagator from the effective four-dimensional operator}
In this section, the GW relation for the effective operator $D_{q}(n,m)$ 
and the properties of the propagator $S(n,m)\equiv (D_{q})^{-1}(n,m)$ 
in the temporal direction are investigated at vanishing spatial momentum and mass.
The GW relation violation is measured using
\begin{align}
\delta(n_4,m_4) = \sum_{\mathrm{color},\mathrm{spin}}\left|
  \{ \gamma_5 , \tilde{D}_q(n_4,m_4)\}
 -2\tilde{D}_q(n_4,m_4)\gamma_5\tilde{D}_q(n_4,m_4)
\right|,
\label{eq:GWVIOLATION}
\end{align}
where $\tilde{D}_q(n_4,m_4)$ is the zero-momentum portion of $D_q(n,m)$.

\begin{figure}[t]
  \begin{center}
  \includegraphics[trim=0 65 0 70,scale=1.0,keepaspectratio,clip]{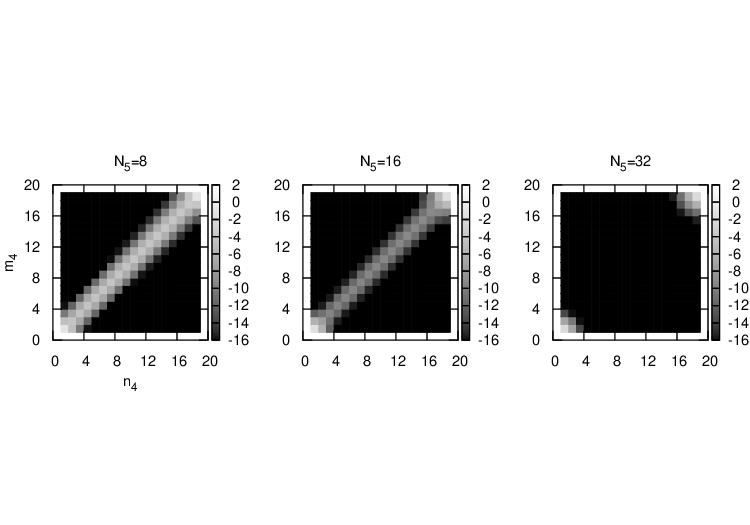}
  \end{center}
  \caption{GW relation violation (\ref{eq:GWVIOLATION}) of the MDWF with $c_j=b_j=w_j$ and $c_{\mathrm{SF}}=1$.
           The $w_j$ are taken from the optimal Zolotarev approximation 
           with the approximation range $x \in [0.01,7.00]$ at the given order. 
           The data are evaluated on the background fields 
(\ref{eq:BF}) with (\ref{eq:BoundaryPhase}) with $\eta=0$.}
  \label{fig:GWVIOLATION}
\end{figure}

Figure~\ref{fig:GWVIOLATION} shows the GW relation violation with the background field, 
$\theta=\pi/5$ and the boundary coefficient $c_{\mathrm{SF}}=1$.
We employ the operator $c_j=b_j=w_j$ applying the quasi-optimal approximation, and 
the approximation range is fixed to $x\in[0.01,7.00]$ for all $N_5$. 
The gray scale corresponds to $\log_{10}(\delta(n_4,m_4))$.
As $N_5$ increases, the violation in the bulk temporal region vanishes 
but remains at the temporal boundary as expected.
The same behavior is observed for other kernel operators 
with different parameters.

The Dirac propagator with the SF boundary condition
has been analytically obtained in the continuum theory~\cite{Sint:1993un, Sint:2010eh}.
Figs~\ref{fig:COMPAREPROP} and \ref{fig:COMPAREPROPT2} show the time dependence of 
the real part of the spin $(1,1)$-component of the Dirac propagator 
using the Dirac representation for the $\gamma$-matrices.
The boundary coefficient is $c_{\mathrm{SF}}=1$ and 
the source times are $m_4=1$ and $2$ in Figs.~\ref{fig:COMPAREPROP} and \ref{fig:COMPAREPROPT2}, respectively.
The solid line represents the analytic solution in the continuum theory. 
The cut-off dependence of the lattice propagator can be compared at
$L/a=N_S=10$ (crosses) and $L/a=N_S=40$ (open circles).

\renewcommand{\figscale}{0.45}
\begin{figure}[t]
  \begin{center}
    \subfigure[The Standard Shamir DWF]{
      \includegraphics[trim=10 5 10 5,scale=\figscale,keepaspectratio,clip]{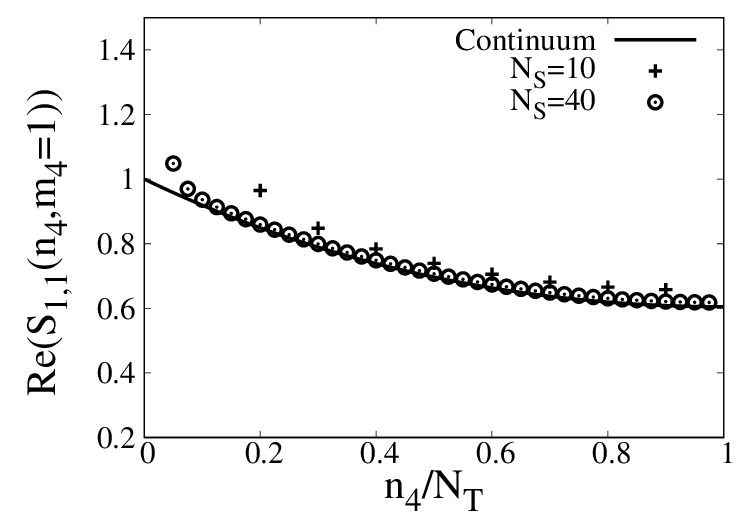}
      \label{fig:PropSdSp11}
    }
    \hfill
    \subfigure[The Bori\c{c}i DWF]{
      \includegraphics[trim=10 5 10 5,scale=\figscale,keepaspectratio,clip]{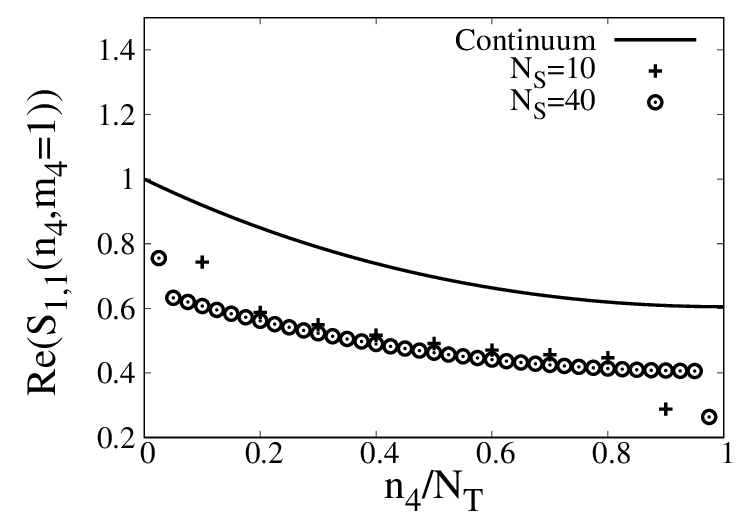}
      \label{fig:PropBdSp11}
    }
  \end{center}
  \caption{Real part of the spin $(1,1)$-component of the Dirac propagator from $m_4=1$. 
The data are evaluated at $am_{res}=0$, 
$am_0=1.0$, $\theta=\pi/5$}, and $N_5=16$ on the zero background 
field ($\phi_k=0$ and $\phi'_k=0$).
  \label{fig:COMPAREPROP}
\end{figure}

\begin{figure}[t]
  \begin{center}
    \subfigure[The Standard Shamir DWF]{
      \includegraphics[trim=10 5 10 5,scale=\figscale,keepaspectratio,clip]{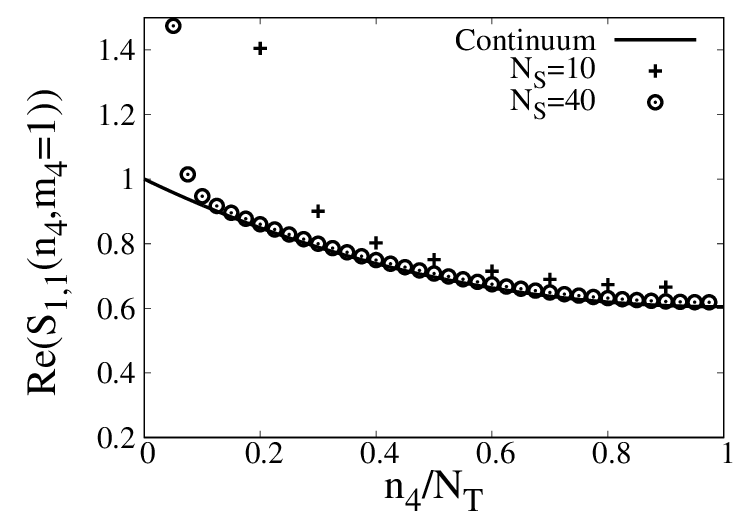}
      \label{fig:PropSdSp11T2}
    }
    \hfill
    \subfigure[The Bori\c{c}i DWF]{
      \includegraphics[trim=10 5 10 5,scale=\figscale,keepaspectratio,clip]{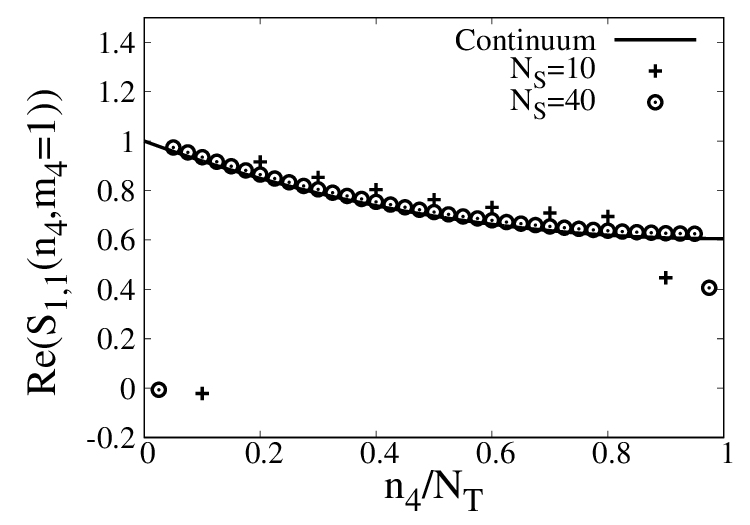}
      \label{fig:PropBdSp11T2}
    }
  \end{center}
  \caption{Real part of the spin $(1, 1)$-component of the Dirac propagator from $m_4=2$.}
  \label{fig:COMPAREPROPT2}
\end{figure}

From the left panel (\ref{fig:PropSdSp11}),
we see that the propagator with the standard domain wall fermion ($b_j=1, c_j=0$)
converges to the continuum limit properly. 
On the other hand, in the right panel (\ref{fig:PropBdSp11})
the propagator with the Bori\c{c}i domain wall 
fermion ($b_j=1,c_j=1$)~\cite{Borici:1999zw,Borici:1999da,Borici:2004pn}
does not converge to the continuum line.
To check the boundary effect further, we plot the Dirac propagator propagating 
from $m_4=2$ in Figure~\ref{fig:COMPAREPROPT2}. 
The left panel (\ref{fig:PropSdSp11T2}) shows the standard domain wall fermion 
and is consistent with the continuum limit.
The Bori\c{c}i domain wall fermion in the right panel (\ref{fig:PropBdSp11T2})
is now consistent with the continuum limit. 
Although the boundary coefficient $c_{\mathrm{SF}}$ can be tuned to eliminate the boundary $O(a)$ error,
it is seen from Figure~\ref{fig:PropBdSp11} that the discrepancy is not proportional to $a$ and cannot be removed by tuning $c_{\mathrm{SF}}$.
We have shown the propagator in the vanishing background field so far, 
the same behavior is seen on the non-zero background field defined by 
Eqs.~(\ref{eq:CFIELD})--(\ref{eq:BoundaryPhase}) with $\eta=0$,
in which the color degeneracy is resolved.

\begin{figure}[t]
  \begin{center}
    \subfigure[Spin $(1,1)$-component.]{
      \includegraphics[scale=\figscale,keepaspectratio,clip]{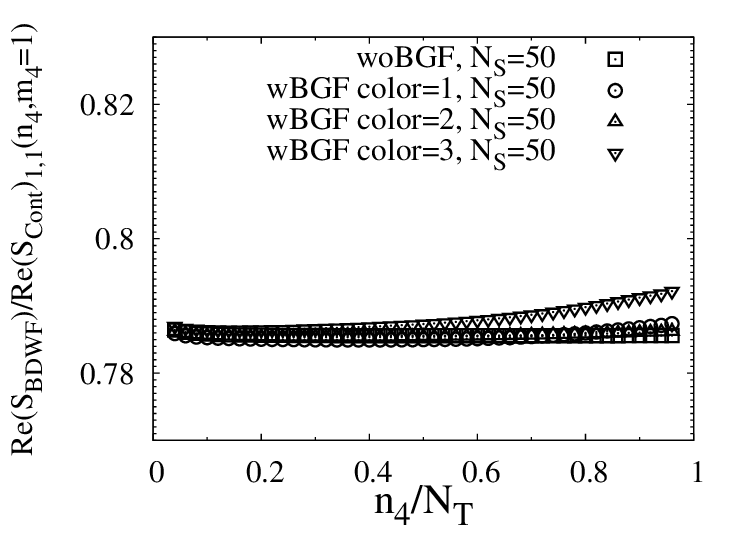}
      \label{fig:PropRatioBdTuneSp11}
    }
    \hfill
    \subfigure[Spin $(3,1)$-component.]{
      \includegraphics[scale=\figscale,keepaspectratio,clip]{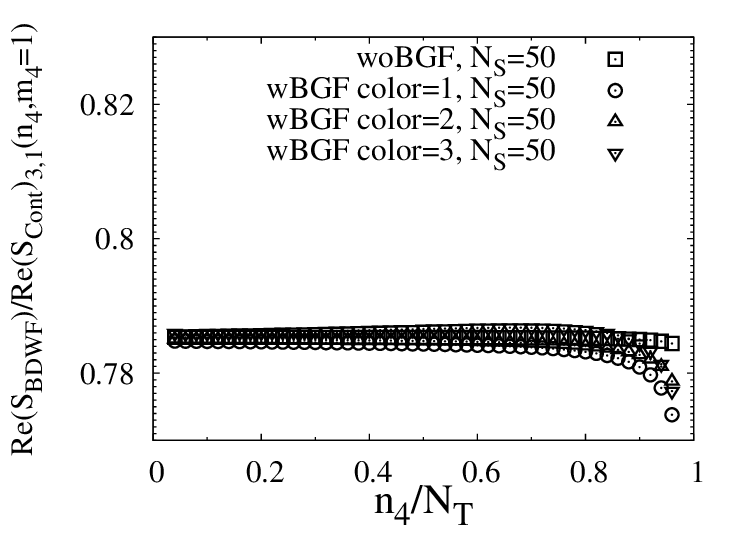}
      \label{fig:PropRatioBdTuneSp31}
    }
  \end{center}
  \caption{The ratio of the real part of the Bori\c{c}i DWF propagator to the continuum theory. The source time is $m_4=1$.
           Squares are for the vanishing background field, while circles, up-, and down-triangles are for the non-zero 
           background field defined by Eqs.~(\ref{eq:CFIELD})--(\ref{eq:BoundaryPhase}) with $\eta=0$.
           The data are evaluated with $c_{\mathrm{SF}}=0.4167$, 
           $am_{res}=0$, $am_0=1.0$, $\theta=\pi/5$, and $N_5=16$.}
  \label{fig:COMPAREPROPTUNET2}
\end{figure}

\begin{figure}[t]
  \centering
  \begin{tabular}{cc}
     \includegraphics[scale=\figscale,keepaspectratio,clip]{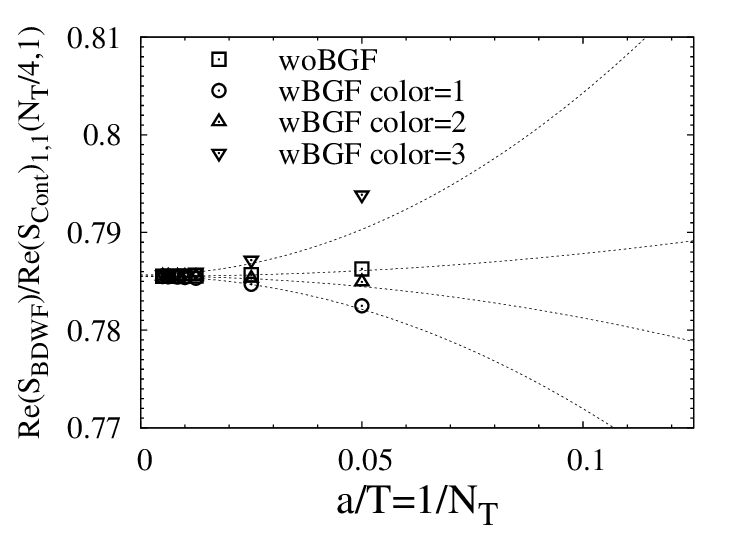} &
     \includegraphics[scale=\figscale,keepaspectratio,clip]{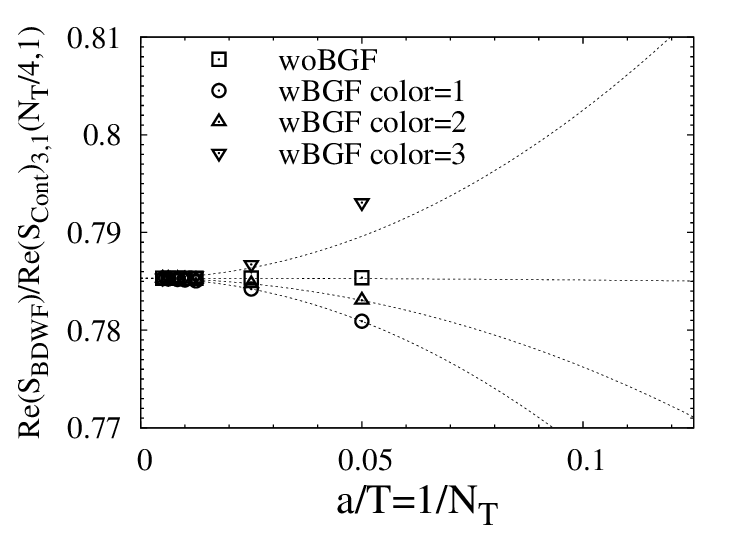} \\
     \includegraphics[scale=\figscale,keepaspectratio,clip]{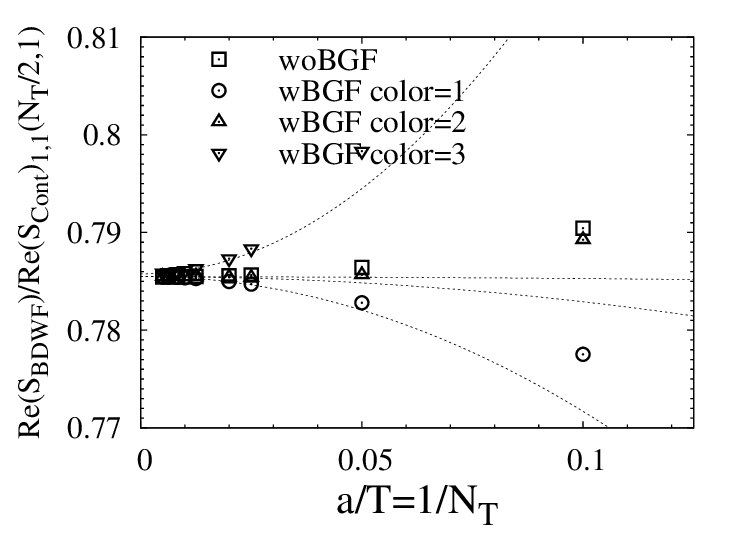} &
     \includegraphics[scale=\figscale,keepaspectratio,clip]{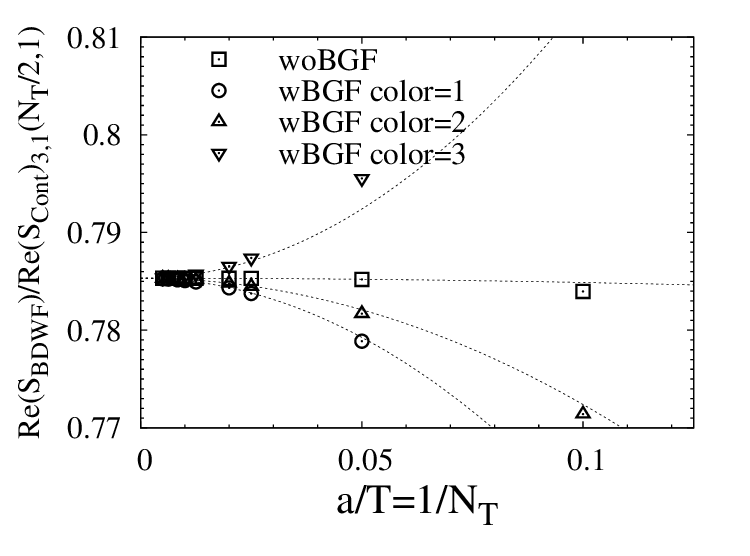} \\
     \includegraphics[scale=\figscale,keepaspectratio,clip]{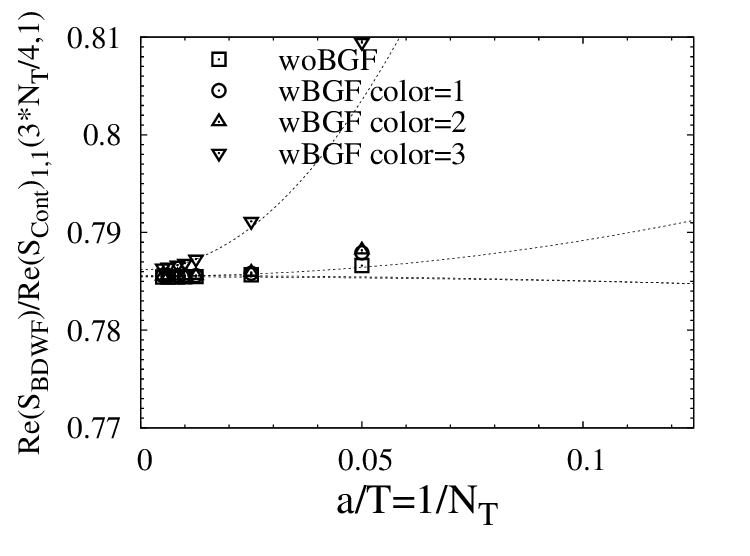} &
     \includegraphics[scale=\figscale,keepaspectratio,clip]{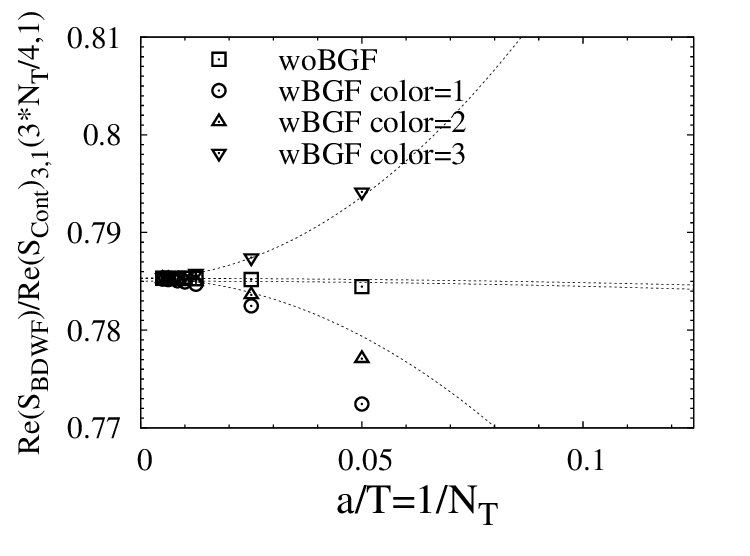} 
  \end{tabular}
  \caption{Continuum limit extrapolation for the ratios of the Bori\c{c}i DWF propagator 
           to the continuum theory (from top to bottom : time slices at $n_4=N_T/4, N_T/2$ and $3 N_T/4$,
           left column : spin $(1,1)$-component,  right column : spin $(3,1)$-component).
           The data are evaluated with $c_{\mathrm{SF}}=0.4167$, 
           $am_{res}=0$, $am_0=1.0$, $\theta=\pi/5$, and $N_5=16$.}
  \label{fig:CONTLIMPropRatio}
\end{figure}

\begin{figure}[t]
  \begin{center}
    \subfigure[Real part of spin $(1,1)$-component of the Dirac propagator in the vanishing background field.]{
      \includegraphics[scale=\figscale,keepaspectratio,clip]{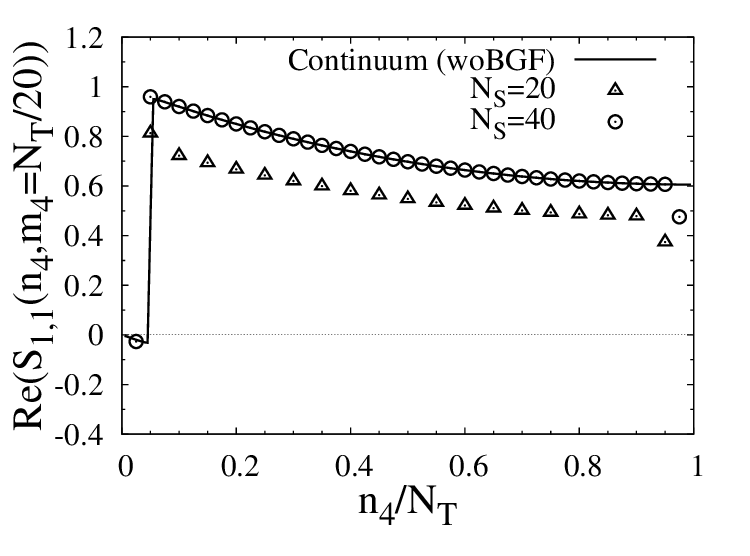}
      \label{fig:PropTsrcNTov20}
    }
    \hfill
    \subfigure[Continuum limit extraporation at the sink time $n_4=N_T/2$.]{
      \includegraphics[scale=\figscale,keepaspectratio,clip]{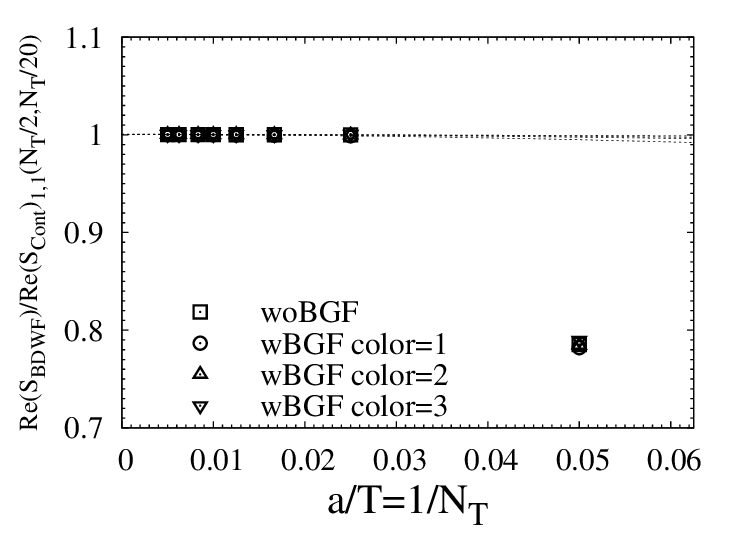}
      \label{fig:PropRatioTsrcNTov20ContLim}
    }
  \end{center}
  \caption{Tho Bori\c{c}i DWF propagator from the source time $m_4=N_T/20$.
           The data are evaluated using the Bori\c{c}i DWF with $c_{\mathrm{SF}}=0.4167$, 
           $am_{res}=0$, $am_0=1.0$, $\theta=\pi/5$, and $N_5=16$.}
  \label{fig:PropCompareTsrcNTov20}
\end{figure}

To better understand the property of the discrepancy,
we investigated the ratio of the propagator of the Bori\c{c}i domain wall fermion to that under continuum theory
and take the continuum limit.
We employ  $c_{\mathrm{SF}}=0.4167$, which is determined by the PCAC relation to be described in the next subsection, 
to eliminate the dominant $O(a)$ error in taking the continuum limit.
We also investigate the dependence of the discrepancy on the presence of the background field.
Figure~\ref{fig:COMPAREPROPTUNET2} shows the time dependence of the ratio of 
the lattice Dirac propagator to the continuum Dirac propagator.
We observe that the discrepancy is almost constant in time and 
slightly depends on the color index.
Figure~\ref{fig:CONTLIMPropRatio} shows the continuum extrapolation for the ratios at the sink time slices at $n_4=N_T/4, N_T/2$, and $3 N_T/4$.
As we employ $c_{\SF}=0.4167$ for the boundary $O(a)$-improvement, the dominant cut-off dependence is of $O(a^2)$.
The dotted lines in the figure show the fitting results with $c + d /N_T^2$ on the data in $< 1/N_T=1/80$ without any constraints 
on the fitting parameters.
The discrepancy converges to a common value (this case $\simeq 0.7856(6)$) irrespective of the spin components, 
sink time,  and the presence of the background field in the continuum limit.
We also obtain the same constant value for the other source time $m_4=N_T-1$. 
Figure~\ref{fig:PropCompareTsrcNTov20} shows 
the propagator from the source time at $m_4=N_T/20$, which is close to the boundary, 
but the physical distance from the boundary is kept fixed at $am_4=y_4=T/20$.
As seen in Figure~\ref{fig:PropTsrcNTov20}, 
the propagator with $N_S=20$ (up-triangles) shows a large deviation from the continuum theory as $m_4=20/20=1$,
while that with $N_S=40$ (circles) almost overlaps on the continuum theory. 
The propagator properly converges to the continuum theory as in Figure~\ref{fig:PropRatioTsrcNTov20ContLim}.

We found that this phenomenon, 
a constant factor discrepancy in the continuum limit remains in 
the propagator at the surface time slices $n_4=1$ and $n_4=N_T-1$ 
(the interior time surfaces of the SF boundary condition)
is common to other MDWFs with $c_j\ne 0$ we investigated.
This strongly suggests the presence of the boundary effect introduced by Eq.~(\ref{eq:BSF}).
As we put the boundary term $B_{SF}$ at $n_4=1$ and $N_T-1$ in the MDWF operator,
contact terms with $B_{SF}$ and the fermion field operators of the effective four-dimensional theory could exist at the surface time slices. 

From these observations, we can conclude that a constant renormalization is needed for the propagator 
touching the surface time slices. 
This phenomenon has been also observed in the overlap fermion in the SF scheme\cite{Luscher:2006df}.
A degree of freedom is available to renormalize the boundary operators or fields
of the fermion fields.\cite{Sint:1995rb,Sint:1993un,Luscher:1996sc,Sint:2010eh}
The boundary fields at $n_4=0$ and $n_4=N_T$ in the SF scheme can be 
renormalized independently from the bulk fermion fields
and the renormalization for the boundary fields has been introduced for the overlap fermion in Ref.~\citen{Luscher:2006df}.
We briefly show the definition of the boundary fields
in the SF scheme\cite{Sint:1995rb,Sint:1993un,Luscher:1996sc,Luscher:2006df}
in the following. After introducing the definition, we discuss two possibilities for 
the renormalization on the surface time slices using the boundary fields.

In the SF scheme, instead of the Dirichlet boundary condition (Eqs.~(\ref{eq:SFBC}) and (\ref{eq:SFBC2})), 
the following inhomogeneous boundary condition,
\begin{align}
      P_{+}\psi(x)|_{x_4=0}=\rho(\bm{x}),\quad P_{-}\psi(x)|_{x_4=T}=\rho'(\bm{x}),
\label{eq:SFBCext}
\\
\bar{\psi}(x)P_{-}|_{x_4=0}=\bar{\rho}(\bm{x}),\quad \bar{\psi}(x)P_{+}|_{x_4=T}=\bar{\rho}'(\bm{x}),
\label{eq:SFBC2ext}
\end{align}
can be imposed on the fermion field in the continuum theory. 
The three-dimensional fields, $\rho(\bm{x})$ and $\rho'(\bm{x})$, 
act as the auxiliary source fields and considered as functional parameters of the partition function 
in the SF scheme\cite{Sint:1993un,Sint:1995rb,Luscher:1996sc}. 
The inhomogeneous boundary condition can not be imposed directly 
on the lattice fields in the bulk temporal region and 
it emerges after taking the continuum limit of the lattice theory provided 
that the lattice action contains proper couplings to the boundary fields $\rho, \rho',\bar{\rho}, \bar{\rho}'$.
According to the method described in Ref.~\citen{Luscher:2006df},
instead of adopting the inhomogeneous boundary condition (Eqs.~(\ref{eq:SFBCext}) and (\ref{eq:SFBC2ext})),
we employ the homogeneous boundary condition
 (Eqs.~(\ref{eq:SFBC}) and (\ref{eq:SFBC2})) and introduce the boundary quark fields directly by
\begin{alignat}{2}
\zeta(\bm{x}) & = P_{-}\psi(x)|_{x_4=0} ,& \qquad 
\zeta'(\bm{x})& = P_{+}\psi(x)|_{x_4=T}
\notag\\
\bar{\zeta}(\bm{x}) & = \bar{\psi}(x)P_{+}|_{x_4=0},& \qquad 
\bar{\zeta}'(\bm{x})& = \bar{\psi}(x)P_{-}|_{x_4=T}
\label{eq:boundaryoperatorscont}
\end{alignat}
in the continuum theory.

A possible form for the boundary fermion fields on the lattice\cite{Luscher:2006df,Takeda:2010ai}
is
\begin{alignat}{2}
\zeta(\bm{n}) & = U_{4}(n-\hat{4})P_{-}\psi(n)|_{n_4=1} ,& \qquad 
\zeta'(\bm{n})& = U_{4}(n)^{\dag} P_{+}\psi(n)|_{n_4=N_T-1}
\notag\\
\bar{\zeta}(\bm{n}) & = \bar{\psi}(n)P_{+}U_{4}(n-\hat{4})^{\dag}|_{n_4=1},& \qquad 
\bar{\zeta}'(\bm{n})& = \bar{\psi}(n)P_{-}U_{4}(n)|_{n_4=N_T-1},
\label{eq:boundaryoperatorslat}
\end{alignat}
where $\psi(n)$ is the effective four-dimensional fermion field in the bulk temporal region 
and $\hat{4}$ is a lattice unit vector in the temporal direction.

Including these boundary fields together with the fermion fields in the bulk time slices,
various fermionic correlation functions (Wick contractions) are introduced to probe the system; 
$
\contraction{}{\zeta}{'(\bm{n})}{\bar{\psi}}
\zeta'(\bm{n})\bar{\psi}(m)
$, 
$
\contraction{}{\zeta}{'(\bm{n})}{\bar{\zeta}}
\zeta'(\bm{n})\bar{\zeta}(\bm{m})
$, 
$
\contraction{}{\psi}{(n)}{\bar{\psi}}
\psi(n)\bar{\psi}(m)
$, $\dots$  \textit{etc}. 
For the PCAC relation, which will be described in the next subsection,
the correlation function 
$
\contraction{}{\psi}{(n)}{\bar{\zeta}}
\psi(n)\bar{\zeta}(\bm{m})
$ is used and Eq.~(\ref{eq:boundaryoperatorslat}) leads to
\begin{align}
\contraction{}{\psi}{(n)}{\bar{\zeta}}
\psi(n)\bar{\zeta}(\bm{m})
 = \left.S(n,m)P_{+}U_4^{\dag}(m-\hat{4})\right|_{m_4=1}.
\label{eq:boundary_bulk_propagator}
\end{align}

Now we discuss two possibilities for 
the renormalization on the surface time slices using the boundary fields.
The boundary-bulk propagator, Eq.~(\ref{eq:boundary_bulk_propagator}), 
shows that the discrepancy in $S(n,m)|_{m_4=1}$ 
can be absorbed into the redefinition of the boundary fields through
\begin{align}
      \zeta_R(\bm{n})  = Z_{\zeta} \zeta(\bm{n}),\quad
      \zeta'_R(\bm{n}) = Z_{\zeta} \zeta'(\bm{n}),\quad
\bar{\zeta}_R(\bm{n})  = Z_{\zeta} \bar{\zeta}(\bm{n}),\quad
\bar{\zeta}'_R(\bm{n}) = Z_{\zeta} \bar{\zeta}'(\bm{n}),
\label{eq:boundary_renormalization}
\end{align}
with a constant $Z_{\zeta}$\cite{Sint:1995rb,Luscher:1996sc,Sint:2010eh}.
This normalization method has been adopted in Ref.~\citen{Luscher:2006df}
for the overlap fermion to recover the canonical normalization for the propagators 
that involve the boundary fields.
One required condition for the renormalization via the boundary fields is 
the localization of the discrepancy at the boundaries.
and the localization requires that the discrepancy should not depend on the global property of the system.
As seen in Figure~\ref{fig:COMPAREPROPTUNET2}, the discrepancy is a constant and 
does not depend on the presence of the background field, 
which shows the independence from the global property.
Figure~\ref{fig:PropTsrcNTov20} also supports the localization of the deficit near the boundary.

Another possibility to remedy this defect is to replace the boundary fermion fields (\ref{eq:boundaryoperatorslat})
to the following extended boundary fields;
\begin{align}
\zeta(\bm{n})       &= U_4(n-2\cdot \hat{4}) U_4(n-\hat{4}) P_{-}\psi(n) |_{n_4=2},
\notag\\
\zeta'(\bm{n})      &= U_4(n+\hat{4})^{\dag} U_4(n)^{\dag} P_{+}\psi(n) |_{n_4=N_T-2},
\notag\\
\bar{\zeta}(\bm{n}) &= \bar{\psi}(n)P_{+}U_4(n-\hat{4})^{\dag} U_4(n-2\cdot \hat{4})^{\dag} |_{n_4=2},
\notag\\
\bar{\zeta}'(\bm{n})&= \bar{\psi}(n)P_{-}U_4(n)                U_4(n + \hat{4})             |_{n_4=N_T-2}.
\label{eq:boundary_fields_v2}
\end{align}
In this case the boundary-bulk correlation function becomes
\begin{align}
\contraction{}{\psi}{(n)}{\bar{\zeta}}
\psi(n)\bar{\zeta}(\bm{m})
 = \left.S(n,m)P_{+}U_4^{\dag}(m-\hat{4})U_4^{\dag}(m-2\cdot\hat{4})\right|_{m_4=2},
\label{eq:boundary_bulk_propagator2}    
\end{align}
by which we exclude the propagators $S$ touching the time surface slices at $n_4=1$ and $N_T-1$.
This is an explicit solution to remove the deficit that we encountered at the tree-level 
as far as the deficit is localized at the boundaries because the renormalization constant $Z_{\zeta}$ is unity in this case.
We will use these boundary to bulk propagators,
Eqs.~(\ref{eq:boundary_bulk_propagator}) and (\ref{eq:boundary_bulk_propagator2}), 
in the next subsection for the $O(a)$-improvement via the PCAC relation.

\subsection{Tuning on the boundary coefficient $c_{\mathrm{SF}}$}

The boundary operator in Eq.~(\ref{eq:BSF}) causes an $O(a)$ error in the spectrum.
Following Refs.~\citen{Luscher:2006df, Takeda:2007ga} and \citen{Takeda:2010ai},
we employed the PCAC relation to remove the error
by tuning the boundary coefficient $c_{\mathrm{SF}}$.

In the continuum, the two-point correlation functions used for 
the PCAC relation are defined by
\begin{align}
    \label{eq:PACACcont}
f_{A}(x_4) &= -\dfrac{1}{(N_f^2-1)L^3}
\sum_{a=1}^{N_f^2-1} \int d^3\bm{x}d^3\bm{y}d^3\bm{z}\langle A_4^a(x)O^a(\bm{y},\bm{z})\rangle,\\
    f_{P}(x_4) &= -\dfrac{1}{(N_f^2-1)L^3}
\sum_{a=1}^{N_f^2 -1} \int d^3\bm{x}d^3\bm{y}d^3\bm{z}\langle P^a(x)O^a(\bm{y},\bm{z})\rangle,\\
A_{\mu}^a(x)&=\bar{\psi}(x)\gamma_\mu\gamma_5T^a\psi(x),\quad
P^a(x) =\bar{\psi}(x)\gamma_5T^a\psi(x),\notag\\
&\qquad
O^a(\bm{y},\bm{z}) =\bar{\zeta}(\bm{y})P_{+} \gamma_5 T^a P_{-}\zeta(\bm{z}),
\label{eq:PSboundary}
\end{align}
where $T^a$ is the generators of $\mathrm{SU}(N_f)$ flavor symmetry.
The two-point functions with vanishing background field in the continuum theory 
at the tree-level become
\begin{align}
 f_A(x_4)&= -\dfrac{N_c}{R^2}  \left[E^2 - m^2 + m \{ m\cosh(2E(T-x_4))+E\sinh(2E(T-x_4)) \} \right],\\
 f_P(x_4)&=  \dfrac{N_c E}{R^2}\left[E \cosh(2E(T-x_4))+ m \sinh(2E(T-x_4))\right],\\
 R&= E\cosh(ET)+m\sinh(ET),\quad
E = \sqrt{\bm{p}_0^2 + m^2},\quad
\bm{p}_0 = (\theta, \theta, \theta)/L, 
\end{align}
where $N_c$ is the number of colors. 
The ratio at $x_4=T/2$ and $T=2L$ with $m=0$ then becomes
\begin{align}
  \dfrac{f_A(T/2)}{f_P(T/2)}= - \dfrac{1}{\cosh(2\sqrt{3}\theta)}.
\label{eq:PCACRAT}
\end{align}

Using the lattice operator $D_q$ and the boundary fields Eq.~(\ref{eq:boundaryoperatorslat}) 
including the renormalization constant $Z_{\zeta}$ via Eq.~(\ref{eq:boundary_renormalization}),
the two-point functions on the lattice are given by
\begin{align}
  f_A(n_4)&= \dfrac{-1}{2N_S^3}\mathrm{Tr}\left[ [\tilde{S}(n_4,1)P_{+}]^{\dag}\gamma_4[\tilde{S}(n_4,1)P_+]\right] Z_{\zeta}^2,
\label{eq:falattice}
\\
  f_P(n_4)&=  \dfrac{1}{2N_S^3}\mathrm{Tr}\left[ [\tilde{S}(n_4,1)P_+]^{\dag}[\tilde{S}(n_4,1)P_+]\right] Z_{\zeta}^2,
\label{eq:fplattice}
\end{align}
where $\tilde{S}(n_4,m_4)$ is the zero-momentum projection of $S(n,m)=(D_q)^{-1}(n,m)$.
At the tree-level, we set $U_4(n)=1$.

The tuning on $c_{\mathrm{SF}}$ is carried out by fitting  $f_A(T/2)/f_P(T/2)$ 
as a polynomial function of $c_{\mathrm{SF}}$ and $a/L$ and 
eliminating the $O(a/L)$ term by tuning $c_{\mathrm{SF}}$.
The boundary field renormalization $Z_{\zeta}$ is not required in the ratio as it cancels out.
We fit $f_A(T/2)/f_P(T/2)$ with
\begin{align}
f_A(T/2)/f_P(T/2) 
 &=        A_{00} 
          + (A_{01} + A_{11}c_{\mathrm{SF}} + A_{21} c_{\mathrm{SF}}^2 + A_{31} c_{\mathrm{SF}}^3)(a/L)   \notag \\
 &\qquad\quad + (A_{02} + A_{12}c_{\mathrm{SF}} + A_{22} c_{\mathrm{SF}}^2 + A_{32} c_{\mathrm{SF}}^3)(a/L)^2 \notag \\
 &\qquad\quad + (A_{03} + A_{13}c_{\mathrm{SF}} + A_{23} c_{\mathrm{SF}}^2 + A_{33} c_{\mathrm{SF}}^3)(a/L)^3 + \cdots.
 \label{eq:EXPDFAFP}
\end{align}
$A_{00}$ can be fixed to its continuum value using 
Eq.~(\ref{eq:PCACRAT}) when there is no background field.
The optimal value of $c_{\mathrm{SF}}$ is thus obtained by solving
\begin{align}
A_{01} + A_{11}c_{\mathrm{SF}} + A_{21} c_{\mathrm{SF}}^2 + A_{31} c_{\mathrm{SF}}^3 = 0.
\end{align}

\begin{table}[t]
\tbl{Optimal values for $c_{\mathrm{SF}}$.}
{\begin{tabular}{lcc|ccc|ccc} \hline\hline
$b_j,c_j$         & $b_j=1,c_j=0$ & $b_j=c_j=1$ &\multicolumn{3}{c|}{$b_j+c_j=w_j,b_j-c_j=1$} & \multicolumn{3}{c}{$b_j=c_j=w_j$}\\ \hline
$m_0$             & $1.5$         & $1.5$       &\multicolumn{3}{c|}{$1.0$}                   & \multicolumn{3}{c}{$1.0$}        \\ \hline
$N_5$             & $32$          & $32$        & $8$     & $16$    & $32$                    & $8$     & $16$    & $32$         \\ \hline
$c_{\mathrm{SF}}$ & $0.520$       & $0.312$     & $0.820$ & $0.630$ & $0.5432$                & $0.553$ & $0.392$ & $0.265$      \\ \hline\hline
\end{tabular}
}
\label{tab:CSFTAB}
\end{table}

Table~\ref{tab:CSFTAB} shows the tuning results for several types of MDWF with 
the tuning performed without the background field.
The quasi-optimal parameters used for $b_j+c_j=w_j,b_j-c_j=1$ (Optimal Shamir) and $b_j=c_j=w_j$ (Optimal Chiu) are
listed in Table~\ref{tab:P11MBPARA} in \ref{appendixTabA}.

To see the effect of the tuning on the $c_{\mathrm{SF}}$, we show
the relative discrepancy of the ratio $f_A(n_4)/f_P(n_4)$ 
between the continuum theory and the lattice theory 
for the Bori\c{c}i domain wall fermion in Figure~\ref{fig:PCACRAT}.
The left figure (\ref{fig:PCACRAT_untuned}) is plotted with $c_{\mathrm{SF}}=1$, while 
the right (\ref{fig:PCACRAT_tuned}) is plotted using the tuned parameter  $c_{\mathrm{SF}}=0.312$.
In both cases, the discrepancy vanishes in the continuum limit ($N_S\to \infty$).
This also shows that the surface field renormalization is simply a constant and 
does not affect the bulk region.
The rate of the convergence is faster for the tuned case (right figure).
To see the convergence rate more explicitly, 
we plot the discrepancy at $t=T/2$ in Figure~\ref{fig:PCACERR} as a function of $1/N_T^2=(a/T)^2$.
With the tuned coefficient, the convergence rate is $O(a^2)$ (Fig.~\ref{fig:PCACRATERR_tuned}).
We also observed a similar behavior for other types of the MDWF, including those with 
the quasi-optimal coefficients for $(b_j, c_j)$ from the Zolotarev approximation.

\renewcommand{\figscale}{0.5}
\begin{figure}[t]
  \begin{center}
    \subfigure[$c_{\mathrm{SF}}=1$.]{
      \includegraphics[trim=10 5 10 25,scale=\figscale,clip]{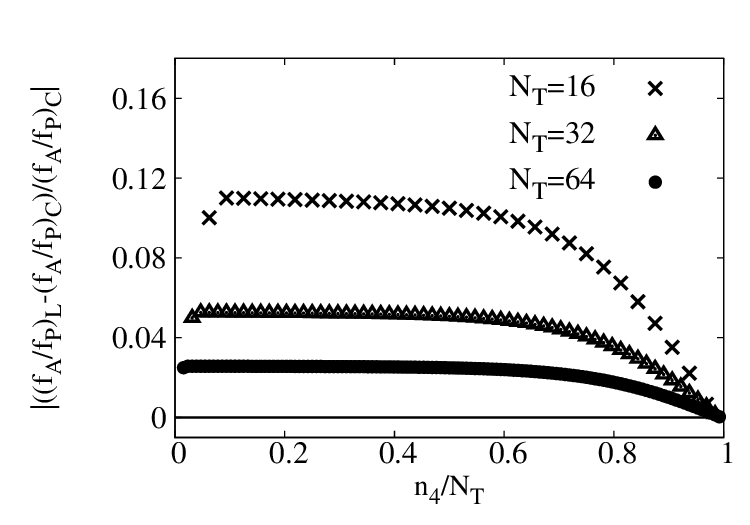}
      \label{fig:PCACRAT_untuned}
    }
\hfill
    \subfigure[$c_{\mathrm{SF}}=0.312$.]{
      \includegraphics[trim=10 5 10 25,scale=\figscale,clip]{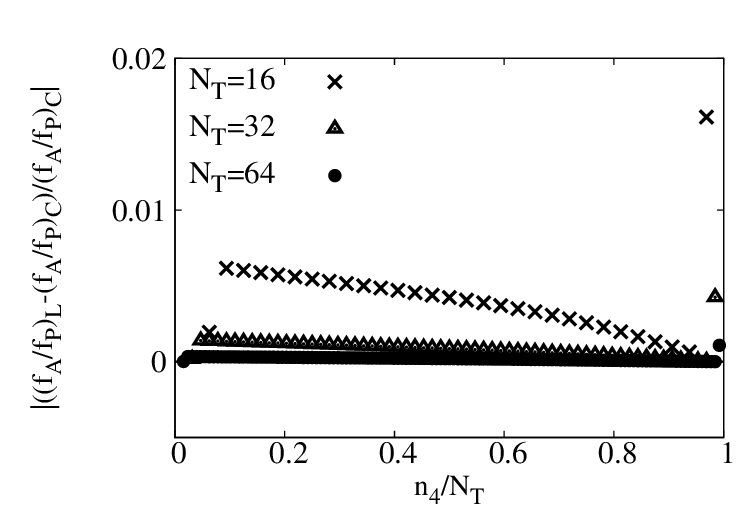}
      \label{fig:PCACRAT_tuned}
    }
  \end{center}
  \caption{Time dependence of the discrepancy of $f_A(n_4)/f_P(n_4)$ 
           between the continuum and lattice theories.
           Bori\c{c}i domain wall fermion ($b_j=c_j=1$) with $N_5=16$ is compared.}
  \label{fig:PCACRAT}
\end{figure}

\begin{figure}[t]
  \begin{center}
    \subfigure[$c_{\SF}=1$.]{
      \includegraphics[trim=10 5 20 25, scale=\figscale,clip]{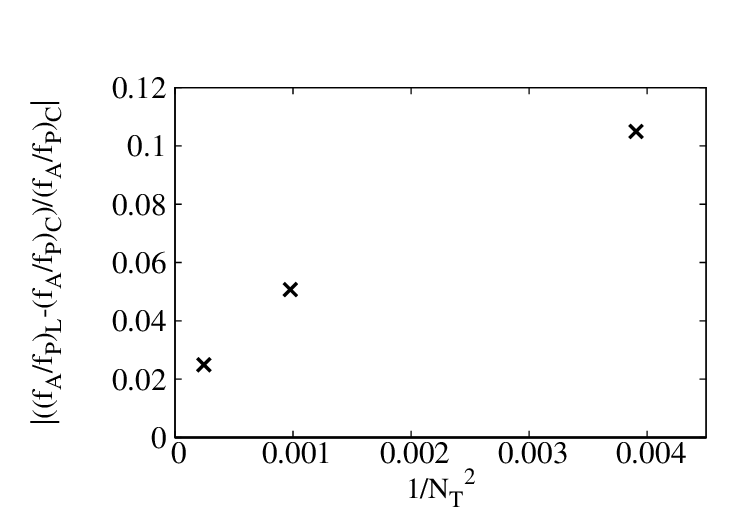}
      \label{fig:PCACRATERR_untuned}
    }
\hfill
    \subfigure[$c_{\mathrm{SF}}=0.312$.]{
      \includegraphics[trim=10 5 20 25, scale=\figscale,clip]{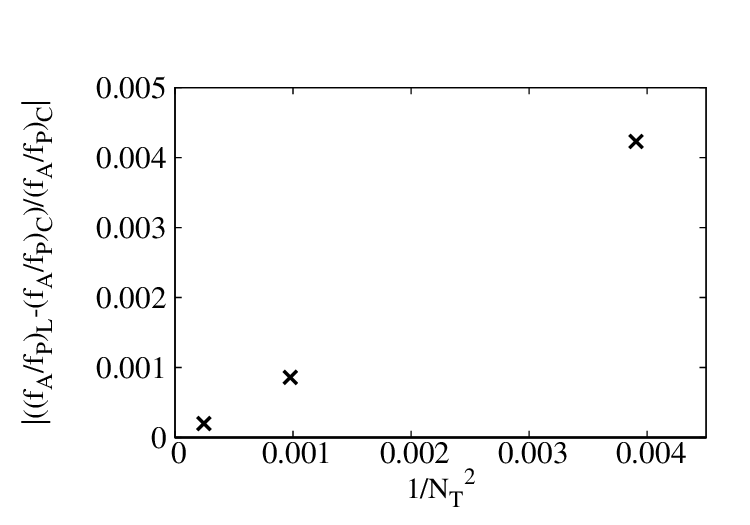}
      \label{fig:PCACRATERR_tuned}
    }
  \end{center}
  \caption{$1/N_T^2$ dependence of the discrepancy of $f_A(N_T/2)/f_P(N_T/2)$ 
           between the continuum and lattice theories.
           Bori\c{c}i domain wall fermion ($b_j=c_j=1$) with $N_5=16$ is compared.}
\label{fig:PCACERR}
\end{figure}

From the observations made in this section,
we can conclude that the MDWF with the SF boundary term in Eq.~(\ref{eq:BSF}) properly reproduces 
the continuum theory in the most bulk regions through the inclusion of the renormalization at 
the tree-level.

Using the boundary fields defined in 
Eq.~(\ref{eq:boundary_fields_v2}) renormalized with Eq.~(\ref{eq:boundary_renormalization}),
we can tune $c_{\mathrm{SF}}$ similarly.  
The two point functions with Eq.~(\ref{eq:boundary_bulk_propagator2}) are
\begin{align}
  f_A(n_4)&= \dfrac{-1}{2N_S^3}\mathrm{Tr}\left[ [\tilde{S}(n_4,2)P_+]^{\dag}\gamma_4[\tilde{S}(n_4,2)P_+]\right]Z_{\zeta}^2,
\label{eq:falatticev2}
\\
  f_P(n_4)&=  \dfrac{1}{2N_S^3}\mathrm{Tr}\left[ [\tilde{S}(n_4,2)P_+]^{\dag}[\tilde{S}(n_4,2)P_+]\right] Z_{\zeta}^2,
\label{eq:fplatticev2}
\end{align}
where the initial time slice of propagators is changed to $n_4=2$ from $n_4=1$ of 
Eqs.~(\ref{eq:falattice}) and~(\ref{eq:fplattice}).
We find that $c_{\mathrm{SF}}$ with Eqs.~(\ref{eq:falatticev2}) and~(\ref{eq:fplatticev2}) is 
almost the same values as listed in Table~\ref{tab:CSFTAB}
and the discrepancies are only in the last digit.

With Eq.~(\ref{eq:boundary_bulk_propagator2}),
there is no the boundary renormalization as $Z_{\zeta}=1$.
We also perform the tuning of $c_{\mathrm{SF}}$ 
using either $f_A(n_4)$ (\ref{eq:falatticev2}) or $f_P(n_4)$ (\ref{eq:fplatticev2}), independently.
This method also yields identical values for $c_{\SF}$ to that obtained with fitting the ratio $f_A/f_P$.

\section{Fermionic contribution to the one-loop beta function}
\label{sec:OneLoopBeta}
The renormalized coupling constant in the SF scheme $g_{\mathrm{SF}}$ is defined by
\begin{align}
\dfrac{1}{g_{\mathrm{SF}}^2} &= \left.\dfrac{1}{k}\dfrac{\partial \Gamma}{\partial\eta}\right|_{\eta=0},
\\
\Gamma &= -\log Z(\eta),
\\
Z(\eta) &= \int{\cal D}U{\cal D}\bar{\psi}{\cal D}\psi
\Psi_{f}[U,\bar{\psi},\psi,\eta]^{*}
\Psi_{i}[U,\bar{\psi},\psi,\eta]
e^{-S_g[U]-S_q[U,\bar{\psi},\psi]},
\end{align}
where
$Z(\eta)$ is the partition function, and 
$\Psi_{i}$ and $ \Psi_{f}$ are the initial ($n_4=0$) and final ($n_4=N_T$) wave-functionals, respectively.
The spatial gauge field at $n_4=0$ and $n_4=N_T$ is fixed 
according to Eq.~(\ref{eq:BF})
by the delta wave-functionals contained in $\Psi_{i}$ and $\Psi_{f}$, respectively.
As seen in Eqs.~(\ref{eq:BF}) and (\ref{eq:BoundaryPhase}), $\eta$ parametrizes the SF boundary condition.
$S_g$ is a lattice gauge action which has Eq.~(\ref{eq:CFIELD}) as the classical solution,
and $S_q$ is a fermion action. 
$k$ is a normalization constant depending on the gauge action $S_g$ and is determined to satisfy
$g_{\mathrm{SF}}=g_0$ at the tree-level.

  We employ the MDWF action defined in Eq.~(\ref{eq:MDWFPVaction}) 
  with the opertor Eq.~(\ref{eq:MDWFSF}) for $S_q$
  by introducing the five-dimensional fermion field $\Psi$ together with the Pauli-Villars field $\Phi$.
Using the coupling expansion and the saddle-point approximation, the effective action $\Gamma$ is
expanded as
\begin{align}
    \Gamma = \dfrac{1}{g_0^2}\Gamma^{(0)} + \Gamma^{(1)} + \cdots.
\end{align}
We focus on the fermionic contribution to $\Gamma^{(1)}$.
The one-loop contribution to $g_{\mathrm{SF}}$ is parametrized as
\begin{align}
g_{\mathrm{SF}}^2 & = g_0^2 + p_1 g_0^4 + \cdots,\quad
p_1 = p_{1,0} + N_f p_{1,1},
\label{eq:GSFcoupuling}
\end{align}
where $N_f$ is the number of flavors included to tag the fermionic contribution $p_{1,1}$. 

The one-loop coefficient, $p_{1,1}$, can be evaluated via
\begin{align}
p_{1,1} &= \dfrac{1}{k}
\sum_{\bm{p}}
\left.
\mathrm{Tr}\left[
\dfrac{\partial \tilde{D}_{\MDWF}^{(N_5)}}{\partial \eta}
\left(\tilde{D}_{\MDWF}^{(N_5)}\right)^{-1}
-
\dfrac{\partial \tilde{D}_{\PV}^{(N_5)}}{\partial \eta}
\left(\tilde{D}_{\PV}^{(N_5)}\right)^{-1}
\right]
\right|_{\eta=0},
\label{eq:SSF}
\end{align}
where the summation on $\bm{p}$ is done using the discrete momenta $\bm{p}=(2\pi \bm{n} + \theta)/N_S, n_k = 0,1,\cdots,N_S-1$.
$\tilde{D}_{\MDWF}^{(N_5)}$ and $\tilde{D}_{\PV}^{(N_5)}$ are the spatial momentum projection of 
$D_{\MDWF}^{(N_5)}$ and $D_{\PV}^{(N_5)}$, respectively.
The trace is over the color, spin, temporal lattice, 
and fifth-direction lattice indices, and 
the asymptotic form in $a \to 0$ is expected to be
\begin{align}
p_{1,1} \sim \sum_{k=0}^{\infty} \left[ r_k + s_k \ln(L/a) \right] (a/L)^k.
\label{eq:ASYMPTF}
\end{align}
To validate our construction of the MDWFs for the SF scheme at the one-loop level, 
we check the following two required conditions:
(i) $s_0$ should coincide with the one-loop beta function $s_0 = 2b_{0,1}=-1/(12\pi^2)\simeq -0.00844343\cdots$; 
(ii) $r_0$ should reproduce the known universal relation of the running coupling constant between 
the SF and $\overline{\mbox{MS}}$ schemes.
The terms $r_1$ and $s_1$ correspond to the $O(a)$ discretization errors; 
whereas $r_1$ can be eliminated by $c_t$, 
the coefficient of the temporal boundary term of the gauge action~\cite{Luscher:1992an, Luscher:1993gh, Takeda:2007ga, Takeda:2010ai, Sint:1995ch}.
$s_1$ must be absorbed by the counter terms in the fermion action. 
When the lattice chiral symmetry in the bulk region is exact, the $O(a)$ error is induced 
only by the temporal boundary effect. In this case, the error can be removed solely by $c_{\SF}$.
At a small $N_5$, where the lattice chiral symmetry in the bulk region is violated, 
another term similar to the clover term in the bulk region is necessary to remove the $O(a)$ error of $s_1$.
Therefore, monitoring $s_1$ provides a test for chiral symmetry.

\begin{table}[t]
  \tbl{The parameter set used for $p_{1,1}$ in Eq.~(\ref{eq:SSF}).}
  {\begin{tabular}{ l|c|c|c|c|c} \hline\hline
    $(b_j,c_j)$ & $m_0$ & $N_5$     & $m_{cr}$        & $c_{\mathrm{SF}}$ & lattice sizes $L/a$ \\ \hline\hline
                &       &  $8$      & $-0.0039062500$ &          & $[4:48]$ \\ \cline{3-4}
        $(1,0)$ & $1.5$ & $16$      & $-0.0000152588$ & $0.520$  & $[4:48]$ \\ \cline{3-4}
                &       & $32$      & $-0.0000000002$ &          & $[4:48]$ \\ \hline
                & $1.0$ & $8,16,32$ & $0.0$           & $0.4167$ & $[4:48]$ \\ \cline{2-5}
        $(1,1)$ & $1.5$ & $8$       & $-0.0000025600$ & $0.312$  & $[4:80]$ \\ \cline{3-4}
                &       & $16,32$   & $0.0$           &          & $[4:48]$ \\ \hline
{$b_j+c_j=w_j$}
                &       & $8$       & $-0.0475081142$ & $0.820$  & $[4:80]$ \\ \cline{3-5}
{$b_j-c_j=1$}
                & $1.0$ & $16$      & $-0.0003037107$ & $0.630$  & $[4:48]$ \\ \cline{3-5}
                &       & $32$      & $-0.0000000216$ & $0.5432$ & $[4:48]$ \\ \hline
                &       & $8$       & $-0.0034031673$ & $0.553$  & $[4:72]$ \\ \cline{3-5}
 $(w_j,w_j)$
								& $1.0$ & $16$      & $-0.0000393319$ & $0.392$  & $[4:72]$ \\ \cline{3-5}
                &       & $32$      & $-0.0000000029$ & $0.265$  & $[4:48]$ \\ 
    \hline\hline
  \end{tabular}}
  \label{tab:PARAP11}
\end{table}

\renewcommand{\figscale}{0.6}
\begin{figure}[t]
	\begin{center}
	\includegraphics[trim=0 0 0 0,scale=\figscale,clip]{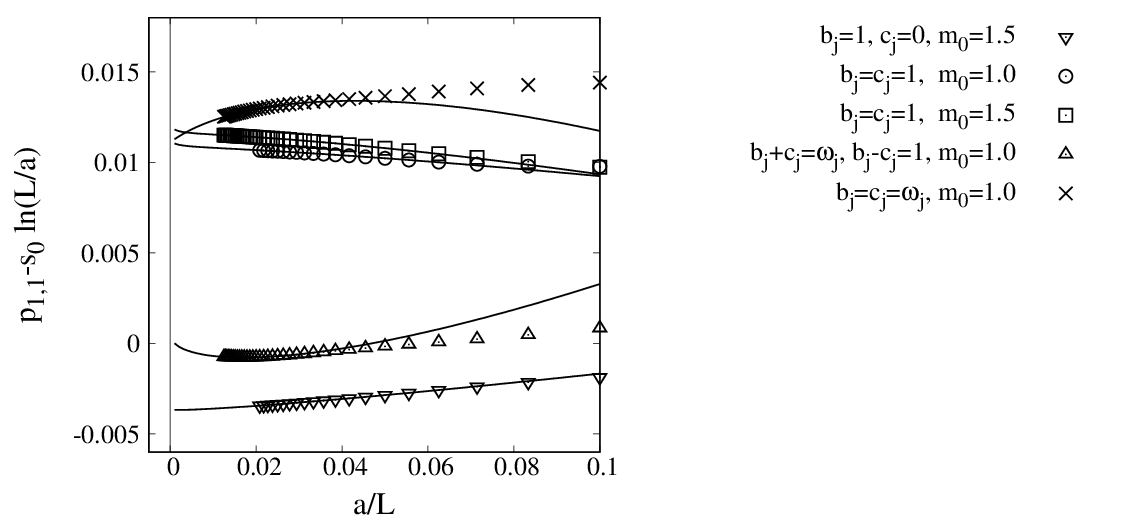}
	\end{center}
	\caption{$p_{1,1}-s_0\ln(L/a)$ as a function of $a/L$ with $N_5=8$.
              The solid lines are the fit results obtained in Table~\ref{tab:P11FREEBESTFIT}.
        }
	\label{fig:FITP11}
\end{figure}

\begin{table}[h]
  \tbl{Best fit results of $p_{1,1}$ with Eq.~(\ref{eq:ASYMPTF}) for each MDWF at $N_5=8$.}
{	\begin{tabular}{l|c|cc|c|c} \hline\hline
       $(b_j,c_j)$ & $b_j=1,c_j=0$ & \multicolumn{2}{c|}{$b_j=c_j=1$} & {$b_j+c_j=w_j$,} & $b_j=c_j=w_j$ \\ 
                   &               &                                & & {$b_j-c_j=1$}    & \\ \hline
	$m_0$ & $1.5$ & $1.0$ & $1.5$ & $1.0$ & $1.0$ \\ \hline
	Fit range & $18 \leq L/a \leq 48$ & $18 \leq L/a \leq 48$ & $16 \leq L/a \leq 80$ & $38 \leq L/a \leq 80$ & $32 \leq L/a \leq72$ \\ \hline
	$r_0$ & $-0.003760(57)$ &  $0.01015(24)$  &  $0.01056(19)$  & $-0.00085(30)$  &  $0.01033(94)$ \\
	$s_0$ & $-0.008430(11)$ & $-0.008318(44)$ & $-0.008268(33)$ & $-0.008299(50)$ & $-0.00835(16)$ \\
	$r_1$ &  $0.03151(41)$  & $-0.0321(17)$   & $-0.0504(18)$   &  $0.1269(55)$   & $-0.132(14)$   \\
	$s_1$ & $-0.00474(31)$  &  $0.0087(13)$   &  $0.0149(12)$   & $-0.0386(28)$   &  $0.0623(76)$  \\ \hline
	$\chi^2$/d.o.f & $1.48 \times 10^{-7}$ & $6.11 \times 10^{-7}$ & $2.67 \times 10^{-6}$ & $3.46\times 10^{-7}$ & $1.40 \times 10^{-6}$ \\
        \hline\hline
	\end{tabular}
	\label{tab:P11FREEBESTFIT}
	}
\end{table}

We numerically evaluated Eq.~(\ref{eq:SSF}) with $N_S=N_T$ using the parameters shown 
in Table~\ref{tab:PARAP11} 
by varying the lattice sizes in steps of two over the given range. 
For the optimal type domain wall fermions, we used the quasi-optimal coefficients $(b_j,c_j)$ shown 
in Table~\ref{tab:P11MBPARA} in \ref{appendixTabA}.
The approximation range was fixed to $x\in [0.001:1.00]$ for the optimal Shamir domain wall fermion 
($b_j+c_j=w_j,b_j-c_j=1$),
and $x\in [0.01:7.00]$ for the optimal Chiu domain wall fermion ($b_j=c_j=w_j$).
As the largest lattice sizes for the optimal type MDWF's satisfy the approximation boundary conditions 
in Eqs.~(\ref{eq:SHAMIR_KERNELRANGE}) and (\ref{eq:OVF_KERNELRANGE}), the continuum limit 
could be taken safely without spoiling the signum function approximation.
All numerically evaluated values for $p_{1,1}$ are tabulated in \ref{appendixTabB}.

We first validated the one-loop beta function $s_0$ by fitting
$p_{1,1}$ as a function of $1/N_S$ assuming the asymptotic form~(\ref{eq:ASYMPTF}) 
including up to $O((a/L)^2)$ terms.
All $r_k$ and $s_k$ were taken as free parameters to validate the one-loop beta function $s_0$.
We varied the cut-off $a/L=1/N_S$ of the fit range $[0:a/L]$ 
and the maximum order of the fitting function
and investigated the stability on the fit result for $s_0$ 
to examine the consistency.
Figure~\ref{fig:FITP11} shows the cut-off $a/L$ dependence of $p_{1,1}+(1/12\pi^2)\ln(L/a)$ with $N_5=8$
as an example of the fitting.
The solid lines are the best fits obtained in the stability analysis and 
corresponding coefficients are listed in Table~\ref{tab:P11FREEBESTFIT}, 
where functions including up to $O(a/L)$ terms yields the best fit.
As seen in the figure and values in the table, 
the optimal    Chiu type ($b_j=c_j=w_j$: crosses) and 
the optimal Shamir type ($b_j+c_j=w_j,b_j-c_j=1$: up-triangles) 
have large values for $s_1$ ($O(a\log(a))$ term),
which indicates that these actions have a less stability on $s_0$ in the fit range analysis.
We observed the stability in an asymptotic region, confirming that 
$s_0$ is consistent with the universal one-loop beta function within 10\% accuracy 
with most of the tested MDWF actions, with the exception of the optimal Chiu type 
($b_j=c_j=w_j$),
for which the discrepancy with $N_5=8$ was 20\%.

We believe that the discrepancy with the optimal Chiu domain wall fermion with $N_5=8$ 
can be explained as follows.
We employed $x\in [0.01:7.0]$ as the approximation range for the quasi-optimal coefficient $b_j$
for the overlap type kernel ${\cal H}_W=\gamma_5D_W$. 
The spectrum of this overlap kernel behaves as in Eq.~(\ref{eq:OVF_KERNELRANGE}), and
limiting the largest lattice size to $N_S=72$. 
This case has a large $O(a)$ error from explicit chiral symmetry breaking and,
therefore, asymptotic behavior to extract the logarithmic cut-off dependence is not captured 
within the narrow fit range.
To extend the fit range towards the asymptotic region in $a/L\to 0$, 
it is necessary to decrease the lower limit of the approximation range. 
For the optimal Chiu MDWF cases with $N_5=16$ and $32$, 
the effect of the explicit chiral symmetry breaking is suppressed, 
and therefore the logarithmic divergence is captured in the fit ranges we examined.

We then investigated $r_0$ in terms of the universal relation of the running coupling constants
between the SF and $\overline{\mbox{MS}}$ schemes.
\footnote{The value of $r_0$ itself depends on the lattice action and 
is not universal because it is computed with lattice regularization in 
the bare coupling expansion. We employ the universal relation of the running coupling constants
between the SF and $\overline{\mbox{MS}}$ schemes to validate $r_0$.}
The one-loop relation between the two schemes is known~\cite{Luscher:1993gh,Sint:1995ch} to be given by
\begin{align}
  \alpha_{\MSbar}(\mu)&=\alpha_{\SF}(\mu)
\left[ 1+ c_1 \alpha_{\SF}(\mu) + \cdots \right], \quad \mu=1/L,
\label{eq:MSbarSF}
\\
c_1 &=  c_{1,0} + N_f c_{1,1},\\
c_{1,0}&=1.25563(4),\quad
c_{1,1}=
\left\{
\begin{matrix}
       0.039863(2) & \qquad \mbox{for}\quad \theta=\pi/5\\
       0.022504(2) & \quad  \mbox{for}\quad \theta=0
\end{matrix}
\label{eq:MSbarSFCoef}
\right.,
\end{align}
where $c_{1,0}$ is the gluonic contribution~\cite{Luscher:1993gh} and 
$c_{1,1}$ is the fermionic contribution~\cite{Sint:1995ch}.
This relation is universal and independent of the regularization used in the SF scheme. 
We employ the lattice regularization with the MDWF in the SF scheme and employ 
the one-loop relation between the lattice bare coupling and the coupling renormalized with the $\SF$ scheme
as in Eq.~(\ref{eq:GSFcoupuling}).
To extract $c_{1,1}$ from $r_0$, it is necessary to know the one-loop relation between 
the lattice bare coupling with the MDWF and the coupling renormalized with the $\MSbar$ scheme.

The MDWF action with $b_j+c_j=w_j,b_j-c_j=1$ (optimal Shamir domain wall fermion) 
is equivalent to the standard domain wall fermion action in the infinite size of $N_5$.
Therefore, $r_0$ is expected to have a common value between the lattice fermions at $N_5=\infty$.
The one-loop relation between the lattice bare coupling with the standard domain wall fermion at $N_5=\infty$ 
and $\alpha_{\MSbar}$ was previously obtained in Ref.~\citen{Aoki:2003uf}; 
by combining these results with the universal relation (\ref{eq:MSbarSF}),
we expect $r_0= 0.0010886(51)$ at $am_0=1.0$ and $r_0=-0.0026362(33)$ at $am_0=1.5$ 
in $N_5\to \infty$ for the optimal Shamir domain wall fermion.

Similarly, the optimal Chiu $(b_j=c_j=w_j)$ and Bori\c{c}i $(b_j=c_j=1)$ DWFs at $N_5=\infty$ 
are equivalent to the overlap fermion action, 
and $r_0$ is expected to have the same value as that derived from the overlap fermion.
In \ref{appendixB}, we show the equivalence of $p_{1,1}$ between the MDWF theory of Eq.~(\ref{eq:MDWFPVaction})
and the truncated overlap fermion theory (\ref{eq:truncatedoverlap}) 
defined with the same operator (\ref{eq:MDWFSF}) at a finite $N_5$ algebraically.
Using the one-loop relation between the lattice bare coupling with the overlap fermion 
and $\alpha_{\MSbar}$ obtained in Ref.~\citen{Alexandrou:1999wr}, we expect 
$r_0=0.01118458(16)$ at $am_0=1.0$ and
$r_0=0.01191070(16)$ at $am_0=1.5$ for the MDWF with $b_j=c_j$ at $N_5=\infty$.

We fit $p_{1,1}$ with Eq.~(\ref{eq:ASYMPTF}) by fixing $s_0=-1/(12\pi^2)$ to extract $r_0$.
We excluded the optimal Chiu DWF with $N_5=8$ from the analysis, 
as we failed the validation on $s_0$ in the fit analysis without constraints.
We employed three fit functions to estimate $r_0$,
\begin{align}
  f(x) &= r_0^{f} + s_0 \ln(x) + (r^{f}_1 + s^{g}_1 \ln(x))/x,\\
  p(x) &= r_0^{p} + s_0 \ln(x) + (r^{p}_1 + s^{p}_1 \ln(x))/x + (r^{p}_2 + s^{p}_2 \ln(x))/x^2,\\
  r(x) &= r_0^{r} + s_0 \ln(x) + (r^{r}_1 + s^{r}_1 \ln(x))/x + (r^{r}_2 + s^{r}_2 \ln(x))/x^2 + (r^{r}_2 + s^{r}_2 \ln(x))/x^3,
\end{align}
where $x=L/a$.
Table~\ref{tab:FITRESULTA0} shows the fit results for $r_0$, in which the error is estimated from 
the stability on the fit results by varying the fit range and changing the fitting function.
We confirm that the values of $r_0$ all agree with the expected values for sufficiently large $N_5$.
From these observations on $s_0$ and $r_0$, 
we conclude that the MDWF with the boundary term in  Eq.~(\ref{eq:BSF}) 
actually satisfies the desired universality at $N_5=\infty$.

To validate $r_0$ at a small $N_5$, 
an independent computation of the one-loop relation 
between the lattice bare coupling and the coupling in the $\MSbar$ scheme 
with the MDWF at the small $N_5$ is required as in 
Refs.~\citen{Aoki:2003uf,Alexandrou:1999wr}.

\begin{table}[t]
  \tbl{Fit results for $r_0$ with $s_0$ is fixed at $-1/(12\pi^2)$.
       The values in the bottom row ($N_5=\infty$) 
       are estimated 
       from the universal relation between the couplings in the SF and $\MSbar$ schemes, 
       and the relation between the coupling in the $\MSbar$ scheme
       and lattice bare coupling scheme
       with the standard domain wall fermion at $N_5=\infty$ and the overlap fermion.}
  {\begin{tabular}{l|c|c|c|c|c} \hline \hline
  $m_0$ & \multicolumn{3}{c|}{$1.0$} & \multicolumn{2}{c}{$1.5$} \\ \hline
  $N_5$ & $b_j+c_j=w_j,$ & $b_j=c_j=1$   &  $b_j=c_j=w_j$  & $b_j=1,c_j=0$    & $b_j=c_j=1$     \\
        & $b_j-c_j=1$    &               &                 &                  &                 \\ \hline \hline
  $8$   & $-0.00100(18)$ & $0.010952(9)$ & --              & $-0.00373594(6)$ & $0.011679(5)$   \\  
  $16$  & $0.001076(15)$ & $0.01116(3)$  & $0.01103(15)$   & $-0.0026696(16)$ & $0.01184(15)$   \\  
  $32$  &  --            & $0.01116(4)$  & $0.0111661(18)$ & $-0.002634(2)$   & $0.01181(4)$    \\ \hline
$\infty$
        & $0.0010886(51)$&\multicolumn{2}{c|}{$0.01118458(16)$}& $-0.0026362(33)$ & $0.01191070(16)$ \\ \hline \hline
  \end{tabular}}
  \label{tab:FITRESULTA0}
\end{table}

\begin{table}[t]
  \tbl{Fit results for $s_1$. $s_0$ is fixed at $-1/(12\pi^2)$ 
       and $r_0$ are fixed at the universal values at $N_5=\infty$.}
  {\begin{tabular}{l|c|c|c|c|c} \hline\hline
  $m_0$ & \multicolumn{3}{c|}{$1.0$} & \multicolumn{2}{c}{$1.5$} \\ \hline
  $N_5$ & $b_j+c_j=w_j,$ & $b_j=c_j=1$   & $b_j=c_j=w_j$ & $b_j=1,c_j=0$ & $b_j=c_j=1$   \\
        & $b_j-c_j=1$    &               &               &               &               \\ \hline \hline
  $16$  & $-0.00115(3)$  & $0.00116(6)$  & $-0.0080(7)$  &  --           & $0.0041(13)$  \\  
  $32$  &  --            & $0.00112(4)$  & $-0.00490(3)$ & $-0.0014(2)$  & $0.0038(9)$   \\
  \hline \hline
  \end{tabular}}
  \label{tab:FITRESULTB1}
\end{table}

We also investigated 
the effect of lattice chiral symmetry on the $O(a)$ error 
and the $O(a)$-improvement through the adjustment of the boundary coefficient $c_{\mathrm{SF}}$.
For this purpose, we evaluated $r_1$ and $s_1$.
After investigating $r_1$ and $s_1$, the lattice artifact on 
the lattice step-scaling function~\cite{Luscher:1991wu} will be discussed.
As mentioned previously, the $O(a)$ error of $r_1$ is absorbed by the boundary counter term
of the gauge action~\cite{Luscher:1996vw,Takeda:2010ai},
while that of $s_1$ can be removed by the boundary coefficient $c_{\mathrm{SF}}$ when the 
lattice chiral symmetry in the bulk region is exact. 
We fit $p_{1,1}$ with
\begin{align}
  f(x) &= r_0 + s_0 \ln(x) + (r^{f}_1 + s^{g}_1 \ln(x))/x,\\
  p(x) &= r_0 + s_0 \ln(x) + (r^{p}_1 + s^{p}_1 \ln(x))/x + (r^{p}_2 + s^{p}_2 \ln(x))/x^2,
\end{align}
where $s_0$ and $r_0$ are fixed to the universal values 
to make the effect of the $O(a)$-improvement by $c_{\SF}$ apparent.
The error on $s_1$ was estimated in the same manner as that of $r_0$. 
Table~\ref{tab:FITRESULTB1} shows the fit results for $s_1$, which would be expected to be zero
when $N_5$ is effectively infinite and $c_{\mathrm{SF}}$ is properly tuned; 
as is seen, the values of $s_1$ are close to, but slightly deviating from, zero.
This deviation can arise from a remaining explicit 
lattice chiral symmetry breaking at finite $N_5$ 
or a miss-tuning of $c_{\mathrm{SF}}$. 
To confirm the $O(a)$-improvement more precisely, more $p_{1,1}$ data at larger $L/a$ 
will be needed to stabilize the fitting to the asymptotic form.

Table~\ref{tab:FITRESULTA1} shows the fit results for $r_1$, where the coefficients $r_0,s_0$ and $s_1$ are fixed 
as $r_0 = \mbox{universal values at $N_5=\infty$}, s_0 = 2b_{0,1}, s_1 = 0$. 
Since actions with $b_j=c_j=w_j,m_0=1.0, N_5=32$ (Optimal Chiu) 
and $b_j=c_j=1,m_0=1.5, N_5=32$ (Borici) 
show slightly larger values for $s_1$ as seen in Table~\ref{tab:FITRESULTB1},
$r_1$ had large errors and could not be determined with the stability analysis.
For other actions, $r_1$ could be determined with regardless of the constraint of $s_1=0$ though large uncertainty remains.

\begin{table}[t]
  \tbl{Fit results for $r_1$. $s_0$ is fixed at $-1/(12\pi^2)$ and $r_0$ are at the universal values at $N_5=\infty$.}
  {\begin{tabular}{l|c|c|c} \hline\hline
  $m_0$      & \multicolumn{2}{c|}{$1.0$}    &  $1.5$         \\ \hline
 $(b_j,c_j)$ & $b_j+c_j=w_j,$ & $b_j=c_j=1$  &  $b_j=1,c_j=0$ \\
             & $b_j-c_j=1$    &              &                \\ \hline\hline
      $N_5$  & $16$           & $32$         &  $32$          \\ \hline
      $r_1$  & $0.014(20)$    & $-0.018(12)$ &  $0.01409(62)$  \\
  \hline \hline
  \end{tabular}}
  \label{tab:FITRESULTA1}
\end{table}

We investigated the lattice cut-off error of the step-scaling function~\cite{Luscher:1991wu} in the SF scheme. 
The step-scaling function $\sigma(2,u)$ with the scaling factor $2$ is defined by
\begin{align}
  \sigma(2,u) = g_{\SF}^2(2L), \quad u=g_{\SF}^2(L).
\end{align}
The lattice version of the step-scaling function at a finite cut-off $a$ is
\begin{align}
  \Sigma(2,u,a/L) = g_{\SF}^2(2L,a), \quad u=g_{\SF}^2(L,a).
\end{align}
The discretization error of the lattice step-scaling function is analyzed through
\begin{align}
  \delta(2,L,a)&\equiv \dfrac{\Sigma(2,u,a/L)-\sigma(2,u)}{\sigma(2,u)}.
\end{align}
The perturbative expansion of the discretization error in terms of the coupling constant $u$ becomes
\begin{align}
  \delta(2,L,a)&= \delta_1(a/L) u + O(u^2),
\end{align}
where the one-loop coefficient $\delta_1(a/L)$ involves pure gauge part and fermionic part~\cite{Sint:1995ch},
\begin{align}
\delta_1(a/L) &= \delta_{1,0}(a/L) + N_f\delta_{1,1}(a/L).
\end{align}
Using Eq.~(\ref{eq:GSFcoupuling}), the discretization error of the fermionic part is given by
\begin{align}
\delta_{1,1}(a/L) &= p_{1,1}(2L/a) - p_{1,1}(L/a) - (2b_{0,1})\log(2).
\label{eq:discerror}
\end{align}
The cut-off dependence of Eq.~(\ref{eq:discerror}) can be parametrized via the asymptotic form of Eq.~(\ref{eq:ASYMPTF}).
The terms with coefficients $r_1$ and $s_1$ in the asymptotic expansion correspond to the $O(a)$ error of the step-scaling function,
and terms with $r_1$ can be removed by absorbing it to $c_t$~\cite{Luscher:1992an, Luscher:1993gh, Takeda:2007ga, Takeda:2010ai, Sint:1995ch}.
We also investigated the $O(a)$-improved version of Eq.~(\ref{eq:discerror}) defined by
\begin{align}
\delta^{(1)}_{1,1}(a/L) &= p_{1,1}(2L/a) - p_{1,1}(L/a) -(2b_{0,1})\log(2)
+ \dfrac{r_1}{2}\left(\dfrac{a}{L}\right),
\label{eq:imporvediscerror}
\end{align}
for actions listed in Table~\ref{tab:FITRESULTA1}.
To clarify the $O(a)$-improvement via $c_t$, we use the symbol $\delta^{(0)}_{1,1}(a/L)= \delta_{1,1}(a/L)$ 
for the unimproved version.

Figure~\ref{fig:SSFBESTFIT0} shows the cut-off $a/L$ dependence of $\delta^{(0)}_{1,1}(a/L)$. 
Wilson and Clover fermions are included for comparison.~\cite{Sint:1995ch}
\footnote{We have reproduced the data and have checked the consistency with that of Ref.~\citen{Sint:1995ch}.}
The Clover fermion does not include the boundary $O(a)$-improvement of removing $r_1$ 
in Fig.~\ref{fig:SSFBESTFIT0}.
Borici type actions (dash-dot and long-dash lines) and
Shamir type action with $am_0=1.5$ (dash line) show non-monotonic behaviors. 
The magnitude of $\delta^{(0)}_{1,1}(a/L)$ is comparable among actions we investigated.
The size is smaller than that of 
Wilson action (three-dashed line) and Clover action (solid line)
 except for Borici action with $m_0=1.5,N_5=32$ (dash-dot line).
Figure~\ref{fig:SSFBESTFIT1} shows the cut-off $a/L$ dependence of $\delta^{(1)}_{1,1}(a/L)$ 
after the $O(a)$-improvement removing $r_1$, where central values in Table~\ref{tab:FITRESULTA1} are used for $r_1$.
Although errors are significantly reduced by the $O(a)$-improvement in the region $a/L < 1/10$, 
errors in $a/L > 1/10$ are larger than that of 
Clover fermion (solid line, $r_1$ is removed)
and Shamir type  with $m_0=1.0, N_5=16$ (dot line) actions. 
It seems that the large errors in $a/L > 1/10$ are caused by the non-monotonic behaviors seen in the unimproved version.

\renewcommand{\figscale}{0.6}
\begin{figure}[t]
  \begin{center}
  \includegraphics[trim=0 0 0 0,scale=\figscale,clip]{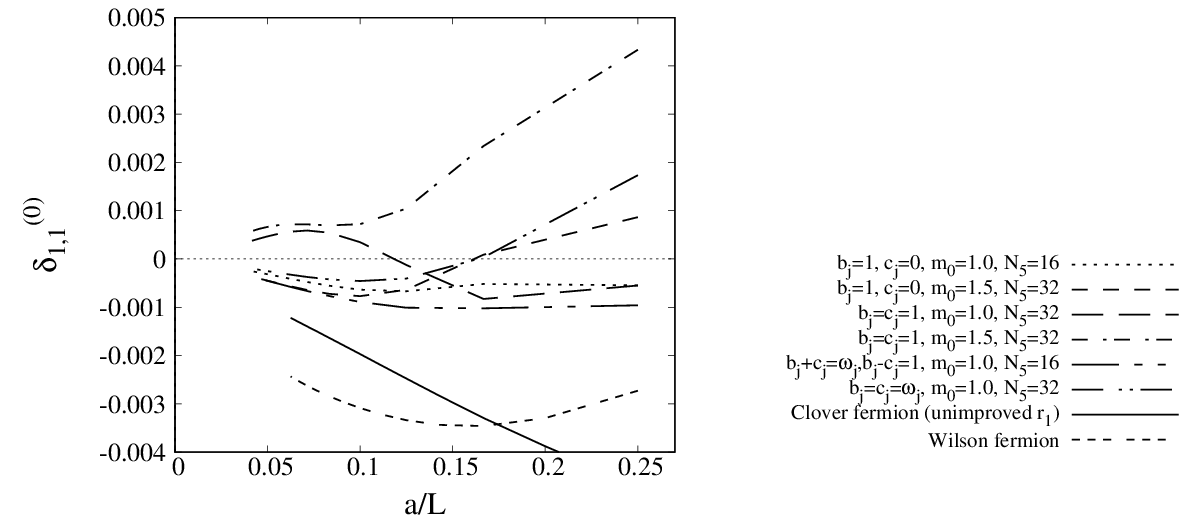}
  \end{center}
  \caption{
    $\delta_{1,1}^{(0)}$ as a function of $a/L$. 
    Dot line ($b_j=1, c_j=0, m_0=1.0, N_5=16$) is taken from Ref.~\citen{Takeda:2010ai} 
    and solid (Clover) and three-dashed (Wilson) lines are reproduced with Ref.~\citen{Sint:1995ch}.
  }
  \label{fig:SSFBESTFIT0}
\end{figure}

\renewcommand{\figscale}{0.6}
\begin{figure}[t]
  \begin{center}
  \includegraphics[trim=0 0 0 0,scale=\figscale,clip]{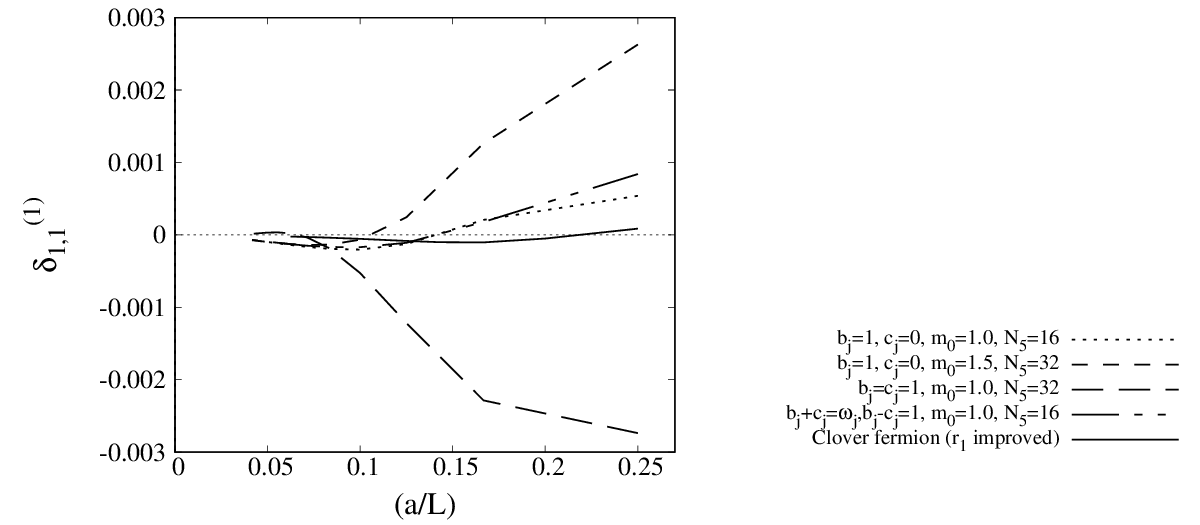}
  \end{center}
  \caption{$\delta_{1,1}^{(1)}$ as a function of $a/L$.}
  \label{fig:SSFBESTFIT1}
\end{figure}

\section{Conclusion}
\label{sec:CONC}
We constructed the appropriate boundary operator for the MDWFs needed
to define the SF scheme by extending Takeda's standard domain wall fermion formulation.

We investigated the effective four-dimensional operator derived from the MDWF at the tree-level and 
confirmed that its spectrum and propagator reproduce the continuum behavior 
in the continuum limit, except for the boundary surfaces,
where the normalization of the propagator differs by a constant from that of the continuum theory.

We proposed two solutions to cure the deficit in the propagator.
One solution is to absorb the discrepancy into the renormalization of the boundary fields
defined in a form usually seen the literature. 
We found that the discrepancy is independent from the presence of the background field,
which satisfies a required condition of the renormalization 
via the renormalization of the boundary fields at the tree-level.
Another solution is to define the boundary fields that are extended 
to two time slices in the bulk time region to avoid the surface time slices of the propagator. 
This works for $f_A$ and $f_P$, and this is the solution to avoid the renormalization.

Using several parameter sets, 
we investigated the fermionic contribution of the MDWF to the one-loop beta function, $p_{1,1}$, 
and were able to extract the one-loop beta function 
consistent with the universal value within a 10\% difference for most of the sets.
The one exception was the case with the optimal Chiu domain wall fermion with $N_5=8$, 
which produced a 20\% deviation from the universal one-loop beta function.
We also examined the universal relation of the couplings between the SF and $\MSbar$ 
and found that 
the cases with sufficiently larger $N_5$ were consistent with the known values from 
the standard domain wall and overlap fermions.
To validate the universality of the coupling relation between the SF and $\MSbar$ 
schemes with the MDWF at smaller $N_5$,
another validation using the relation between the $\MSbar$ and bare lattice couplings with the MDWF will be required.
We also investigated the lattice artifact of the lattice step-scaling function at the one-loop level,
and found that the size of the lattice artifact is comparable to that of the standard domain wall fermion.

Based on these results, we can conclude that the boundary operator we introduced is one of the realizations 
of the SF scheme for the  M\"{o}bius domain wall fermions, at least at larger $N_5$ or $N_5=\infty$.

\section*{Acknowledgments}

We would like to thank S.~Takeda for the computational advice and the useful discussion. 
We thank anonymous referee for his/her helpful comments on the surface field renormalization.
The numerical computations have been done on the workstations of the
INSAM (Institute for Nonlinear Sciences and Applied Mathematics) at Hiroshima University 
and on FUJITSU PRIMERGY CX400 (Tatara system) and new FUJITSU PRIMERGY (ITO system)
of the Research Institute for Information Technology at Kyushu University.
This work was partly supported by JSPS KAKENHI Grant Numbers 24540276 and 16K05326, and
Core of Research for the Energetic Universe (CORE--U) at Hiroshima University.

\clearpage
\appendix

\section{Tables for quasi-optimal parameters $b_j$}
\label{appendixTabA}
\setcounter{table}{0}
\renewcommand{\thetable}{A\arabic{table}}

In this appendix, we list the quasi-optimal parameters $b_j$ for the $N_5=8, 16$ and $32$ cases.
The approximation range is fixed to 
$x \in [0.001:1.00]$ for the optimal Shamir domain wall fermion (OSDWF, $b_j+c_j=w_j,b_j-c_j=1$) 
and 
$x \in [0.01:7.00]$  for the optimal Chiu domain wall fermion (OCDWF, $b_j=c_j=w_j$).
We introduce $\delta^{N_5} = |1-R_{N_5}|$ to see the quality of the approximation.

\begin{table}[H]
    \tbl{The MDWF $b_j$ parameters.}
    {\begin{tabular}{lccc} \toprule
                   \multicolumn{4}{c}{OSDWF}                      \\ \colrule
$N_5$ &        $8$        &       $16$        &      $32$         \\ \colrule
$\delta^{N_5}$ 
      &   $\leq 0.091$    &  $\leq 0.00061$    & $\leq 4.3 \times 10^{-8} $  \\ \colrule
  $j$ &  \multicolumn{3}{c}{$b_j$}                                \\ \colrule
  $1$ &  $1.375560445952$ &  $1.070238032009$ & $0.016892137272$  \\
  $2$ &  $5.471160816064$ &  $1.730342867161$ & $0.158913006904$  \\
  $3$ &  $50.79006488628$ &  $3.876257791467$ & $0.482041690426$  \\
  $4$ &  $286.0314000944$ &  $9.868218005280$ & $1.074927835108$  \\
  $5$ &                   &  $27.18597163933$ & $2.100739142300$  \\
  $6$ &                   &  $74.54647850997$ & $3.840873748406$  \\
  $7$ &                   &  $203.6953910351$ & $6.774084261660$  \\
  $8$ &                   &  $438.9134097812$ & $11.70429696596$  \\
  $9$ &                   &                   & $19.98458839515$  \\
 $10$ &                   &                   & $33.86858730352$  \\
 $11$ &                   &                   & $57.09209193583$  \\
 $12$ &                   &                   & $95.62651877838$  \\
 $13$ &                   &                   & $158.2374319172$  \\
 $14$ &                   &                   & $254.0716769840$  \\
 $15$ &                   &                   & $378.9127561311$  \\
 $16$ &                   &                   & $483.1599011152$  \\ \botrule \\ \toprule
                   \multicolumn{4}{c}{OCDWF}                      \\ \colrule
$N_5$ &        $8$        &      $16$         &      $32$         \\ \colrule
$\delta^{N_5}$ 
      &   $\leq 0.072$    &  $\leq 0.00039$    & $\leq 1.8 \times 10^{-8}$   \\ \colrule
  $j$ &  \multicolumn{3}{c}{$b_j$}                                \\ \colrule
  $1$ &  $0.237921110807$ &  $0.161123304598$ &  $0.147274921719$ \\
  $2$ &  $1.265028710883$ &  $0.331200294113$ &  $0.184266408796$ \\
  $3$ & $11.292798466011$ &  $0.862987775592$ &  $0.267551476942$ \\
  $4$ & $60.043912191189$ &  $2.293352637899$ &  $0.418026489555$ \\
  $5$ &                   &  $6.229183445073$ &  $0.673528275619$ \\
  $6$ &                   & $16.553785221249$ &  $1.098151069375$ \\
  $7$ &                   & $43.133157003882$ &  $1.798613861602$ \\
  $8$ &                   & $88.663240375751$ &  $2.950400432880$ \\
  $9$ &                   &                   &  $4.841957764956$ \\
 $10$ &                   &                   &  $7.942624367962$ \\
 $11$ &                   &                   & $13.008878909391$ \\
 $12$ &                   &                   & $21.210266595825$ \\
 $13$ &                   &                   & $34.174184274611$ \\
 $14$ &                   &                   & $53.394264344831$ \\
 $15$ &                   &                   & $77.527501507451$ \\
 $16$ &                   &                   & $97.000318309165$ \\ \botrule
    \end{tabular}
    \label{tab:P11MBPARA}
    }
\end{table}

\clearpage
\section{Tables for $p_{1,1}$}
\label{appendixTabB}
\setcounter{table}{0}
\renewcommand{\thetable}{B\arabic{table}}

In this appendix, we provide tables for the numerical values of $p_{1,1}$.

\begin{table}[H]
  \tbl{The numerical values of $p_{1,1}$ with $b_j+c_j=w_j$, $b_j-c_j=1$, and $m_0=1.0$.}
  {\begin{tabular}{lcc} \toprule
	$N_5$ & $8$ & $16$ \\ \colrule
	$L/a$& \multicolumn{2}{c}{$p_{1,1}$} \\ \colrule
   $4$ & $-0.008078678483014$ & $-0.007549846131326$ \\ 
   $6$ & $-0.012704626284028$ & $-0.011543790007817$ \\ 
   $8$ & $-0.016099359107337$ & $-0.014362647277993$ \\ 
  $10$ & $-0.018603053000762$ & $-0.016601560471003$ \\ 
  $12$ & $-0.020509350351466$ & $-0.018419096227329$ \\ 
  $14$ & $-0.022043043374366$ & $-0.019932682101795$ \\ 
  $16$ & $-0.023334785689534$ & $-0.021224758101492$ \\ 
  $18$ & $-0.024458027530347$ & $-0.022349705318640$ \\ 
  $20$ & $-0.025455706039567$ & $-0.023344258955488$ \\ 
  $22$ & $-0.026354442980561$ & $-0.024234463201614$ \\ 
  $24$ & $-0.027171888915236$ & $-0.025039552752149$ \\ 
  $26$ & $-0.027920587948598$ & $-0.025774134656849$ \\ 
  $28$ & $-0.028610040120546$ & $-0.026449513032584$ \\ 
  $30$ & $-0.029247813609523$ & $-0.027074562214969$ \\ 
  $32$ & $-0.029840155519369$ & $-0.027656336722746$ \\ 
  $34$ & $-0.030392340093947$ & $-0.028200511774470$ \\ 
  $36$ & $-0.030908880213565$ & $-0.028711707029837$ \\ 
  $38$ & $-0.031393667187825$ & $-0.029193726826378$ \\ 
  $40$ & $-0.031850071712519$ & $-0.029649739643784$ \\ 
  $42$ & $-0.032281022324532$ & $-0.030082412981785$ \\ 
  $44$ & $-0.032689069460135$ & $-0.030494015386321$ \\ 
  $46$ & $-0.033076439227606$ & $-0.030886494192633$ \\ 
  $48$ & $-0.033445079148571$ & $-0.031261535256350$ \\ 
  $50$ & $-0.033796697257952$ & \\ 
  $52$ & $-0.034132795554748$ & \\ 
  $54$ & $-0.034454698597199$ & \\ 
  $56$ & $-0.034763577918903$ & \\ 
  $58$ & $-0.035060472857527$ & \\ 
  $60$ & $-0.035346308315376$ & \\ 
  $62$ & $-0.035621909905414$ & \\ 
  $64$ & $-0.035888016875790$ & \\ 
  $66$ & $-0.036145293150457$ & \\ 
  $68$ & $-0.036394336774732$ & \\ 
  $70$ & $-0.036635688010451$ & \\ 
  $72$ & $-0.036869836288573$ & \\ 
  $74$ & $-0.037097226193355$ & \\ 
  $76$ & $-0.037318262626538$ & \\ 
  $78$ & $-0.037533315274019$ & \\ 
  $80$ & $-0.037742722480864$ & \\
  \botrule
  \end{tabular}
  \label{tab:FITP11Ps}
  }
\end{table}

\begin{table}[H]
	\begin{center}
	\tbl{The numerical values of $p_{1,1}$ with $b_j=c_j=w_j$ and $m_0=1.0$.}
	{
	\begin{tabular}{lccc} \toprule
	$N_5$ & $8$ & $16$ & $32$ \\ \colrule
	$L/a$& \multicolumn{3}{c}{$p_{1,1}$} \\ \colrule
   $4$ & $-0.000427658644074$ & $-0.000383843417187$ & $-0.001367108292250$ \\ 
   $6$ & $-0.001690115887320$ & $-0.002015195493540$ & $-0.003292843399314$ \\ 
   $8$ & $-0.003276057963329$ & $-0.004212644775318$ & $-0.005485336972469$ \\ 
  $10$ & $-0.005033155003610$ & $-0.006246918403858$ & $-0.007449843475257$ \\ 
  $12$ & $-0.006705005271095$ & $-0.007992466113233$ & $-0.009112723836382$ \\ 
  $14$ & $-0.008193831514685$ & $-0.009490189069814$ & $-0.010524786720790$ \\ 
  $16$ & $-0.009494057300111$ & $-0.010790678620702$ & $-0.011742834419345$ \\ 
  $18$ & $-0.010630559779250$ & $-0.011933438152156$ & $-0.012810419169036$ \\ 
  $20$ & $-0.011632825706636$ & $-0.012948842121387$ & $-0.013759177135561$ \\ 
  $22$ & $-0.012526885001577$ & $-0.013860522358465$ & $-0.014612217134086$ \\ 
  $24$ & $-0.013333735423218$ & $-0.014686861614616$ & $-0.015386734394376$ \\ 
  $26$ & $-0.014069766215183$ & $-0.015442121609423$ & $-0.016095786578485$ \\ 
  $28$ & $-0.014747628848984$ & $-0.016137387857025$ & $-0.016749483310047$ \\ 
  $30$ & $-0.015377075017193$ & $-0.016781345974691$ & $-0.017355791924810$ \\ 
  $32$ & $-0.015965645179162$ & $-0.017380889556003$ & $-0.017921092828360$ \\ 
  $34$ & $-0.016519204136393$ & $-0.017941576933597$ & $-0.018450569412746$ \\ 
  $36$ & $-0.017042349063875$ & $-0.018467964970409$ & $-0.018948486893592$ \\ 
  $38$ & $-0.017538717989105$ & $-0.018963849305390$ & $-0.019418395305982$ \\ 
  $40$ & $-0.018011222352760$ & $-0.019432436525828$ & $-0.019863279824302$ \\ 
  $42$ & $-0.018462222102058$ & $-0.019876468167950$ & $-0.020285673857459$ \\ 
  $44$ & $-0.018893657364280$ & $-0.020298311160125$ & $-0.020687745384951$ \\ 
  $46$ & $-0.019307147334419$ & $-0.020700025025448$ & $-0.021071363741383$ \\ 
  $48$ & $-0.019704064430391$ & $-0.021083412945066$ & $-0.021438151897327$ \\ 
  $50$ & $-0.020085589825083$ & $-0.021450061496227$ & \\ 
  $52$ & $-0.020452754991647$ & $-0.021801372306609$ & \\ 
  $54$ & $-0.020806472777243$ & $-0.022138587806451$ & \\ 
  $56$ & $-0.021147560664753$ & $-0.022462812558580$ & \\ 
  $58$ & $-0.021476758226712$ & $-0.022775031183158$ & \\ 
  $60$ & $-0.021794740276988$ & $-0.023076123590416$ & \\ 
  $62$ & $-0.022102126845674$ & $-0.023366878034668$ & \\ 
  $64$ & $-0.022399490814134$ & $-0.023648002369445$ & \\ 
  $66$ & $-0.022687363830450$ & $-0.023920133794130$ & \\ 
  $68$ & $-0.022966240962226$ & $-0.024183847319741$ & \\ 
  $70$ & $-0.023236584421461$ & $-0.024439663137563$ & \\ 
	\botrule
	\end{tabular}
  \label{tab:FITP11Pc}
	}
	\end{center}
\end{table}

\begin{table}[H]
	\begin{center}
	\tbl{The numerical values of $p_{1,1}$ with $b_j=1, c_j=0$ and $m_0=1.5$.}
	{
	\begin{tabular}{lccc} \toprule
	$N_5$ & $8$ & $16$ & $32$ \\ \colrule
	$L/a$& \multicolumn{3}{c}{$p_{1,1}$} \\ \colrule
   $4$ & $-0.013700543988949$ & $-0.013494117365375$ & $-0.013493847599046$ \\ 
   $6$ & $-0.016940425660898$ & $-0.016440093560490$ & $-0.016435046256293$ \\ 
   $8$ & $-0.019207595711335$ & $-0.018495725966135$ & $-0.018480529959025$ \\ 
  $10$ & $-0.021308125813259$ & $-0.020479834297536$ & $-0.020459083387075$ \\ 
  $12$ & $-0.023122800312741$ & $-0.022221285948861$ & $-0.022198883266909$ \\ 
  $14$ & $-0.024665375229509$ & $-0.023711578334847$ & $-0.023688561490842$ \\ 
  $16$ & $-0.025987047538505$ & $-0.024995171537818$ & $-0.024971355383199$ \\ 
  $18$ & $-0.027135213800013$ & $-0.026116105185960$ & $-0.026091198129590$ \\ 
  $20$ & $-0.028146880041840$ & $-0.027108518882633$ & $-0.027082437011268$ \\ 
  $22$ & $-0.029049775493944$ & $-0.027997747140651$ & $-0.027970577120069$ \\ 
  $24$ & $-0.029864559491185$ & $-0.028802679063304$ & $-0.028774578712134$ \\ 
  $26$ & $-0.030606757871299$ & $-0.029537633821171$ & $-0.029508765297232$ \\ 
  $28$ & $-0.031288217026821$ & $-0.030213664582947$ & $-0.030184166135890$ \\ 
  $30$ & $-0.031918135064723$ & $-0.030839449899792$ & $-0.030809430378576$ \\ 
  $32$ & $-0.032503778953037$ & $-0.031421910597608$ & $-0.031391453294570$ \\ 
  $34$ & $-0.033050983428172$ & $-0.031966644411249$ & $-0.031935813074617$ \\ 
  $36$ & $-0.033564501913902$ & $-0.032478237562439$ & $-0.032447081752874$ \\ 
  $38$ & $-0.034048257349549$ & $-0.032960491706605$ & $-0.032929050814620$ \\ 
  $40$ & $-0.034505524685294$ & $-0.033416591708052$ & $-0.033384897755921$ \\ 
  $42$ & $-0.034939066008357$ & $-0.033849231429220$ & $-0.033817310976307$ \\ 
  $44$ & $-0.035351232253616$ & $-0.034260709315154$ & $-0.034228584756533$ \\ 
  $46$ & $-0.035744040922304$ & $-0.034653001956654$ & $-0.034620692426401$ \\ 
  $48$ & $-0.036119236284169$ & $-0.035027821389594$ & $-0.034995343407687$ \\ 
	\botrule
	\end{tabular}
  \label{tab:FITP11Sdm015}
	}
	\end{center}
\end{table}

\begin{table}[H]
	\begin{center}
	\tbl{The numerical values of $p_{1,1}$ with $b_j=c_j=1$ and $m_0=1.5$.}
	{
	\begin{tabular}{lccc} \toprule
	$N_5$ & $8$ & $16$ & $32$ \\ \colrule
	$L/a$& \multicolumn{3}{c}{$p_{1,1}$} \\ \colrule
   $4$ & $-0.006727819825856$ & $-0.006665138308415$ & $-0.006666313597976$ \\ 
   $6$ & $-0.007414851979212$ & $-0.007313811347951$ & $-0.007315034976460$ \\ 
   $8$ & $-0.008380305729795$ & $-0.008194515940394$ & $-0.008185897962635$ \\ 
  $10$ & $-0.009706986889307$ & $-0.009534951147513$ & $-0.009513426996691$ \\ 
  $12$ & $-0.010940734313774$ & $-0.010849913953446$ & $-0.010828496014809$ \\ 
  $14$ & $-0.012005880912213$ & $-0.012000303082120$ & $-0.011989158728680$ \\ 
  $16$ & $-0.012933090094362$ & $-0.012990681811199$ & $-0.012992846763797$ \\ 
  $18$ & $-0.013759688673405$ & $-0.013854127110319$ & $-0.013868798328790$ \\ 
  $20$ & $-0.014511760217989$ & $-0.014620713721791$ & $-0.014645831748636$ \\ 
  $22$ & $-0.015205517755086$ & $-0.015312808751915$ & $-0.015346126770424$ \\ 
  $24$ & $-0.015851036926812$ & $-0.015946323963520$ & $-0.015985730259767$ \\ 
  $26$ & $-0.016454984507842$ & $-0.016532563270011$ & $-0.016576163504983$ \\ 
  $28$ & $-0.017022194974289$ & $-0.017079658119618$ & $-0.017125805557620$ \\ 
  $30$ & $-0.017556489774245$ & $-0.017593563907684$ & $-0.017640879135091$ \\ 
  $32$ & $-0.018061078280186$ & $-0.018078737359153$ & $-0.018126116251055$ \\ 
  $34$ & $-0.018538749124392$ & $-0.018538597095271$ & $-0.018585202424702$ \\ 
  $36$ & $-0.018991964014254$ & $-0.018975837134752$ & $-0.019021074631616$ \\ 
  $38$ & $-0.019422909145805$ & $-0.019392639865905$ & $-0.019436123561516$ \\ 
  $40$ & $-0.019833529265709$ & $-0.019790820185260$ & $-0.019832332961955$ \\ 
  $42$ & $-0.020225554783800$ & $-0.020171922778777$ & $-0.020211377175424$ \\ 
  $44$ & $-0.020600525735608$ & $-0.020537287838630$ & $-0.020574690536960$ \\ 
  $46$ & $-0.020959813694587$ & $-0.020888095769654$ & $-0.020923517571072$ \\ 
  $48$ & $-0.021304641794774$ & $-0.021225398065531$ & $-0.021258949911228$ \\ 
  $50$ & $-0.021636102804375$ & & \\ 
  $52$ & $-0.021955175221099$ & & \\ 
  $54$ & $-0.022262737442725$ & & \\ 
  $56$ & $-0.022559580138765$ & & \\ 
  $58$ & $-0.022846416992679$ & & \\ 
  $60$ & $-0.023123893996791$ & & \\ 
  $62$ & $-0.023392597486309$ & & \\ 
  $64$ & $-0.023653061081057$ & & \\ 
  $66$ & $-0.023905771690838$ & & \\ 
  $68$ & $-0.024151174719892$ & & \\ 
  $70$ & $-0.024389678584948$ & & \\ 
  $72$ & $-0.024621658651276$ & & \\ 
  $74$ & $-0.024847460666186$ & & \\ 
  $76$ & $-0.025067403767381$ & & \\ 
  $78$ & $-0.025281783121781$ & & \\ 
  $80$ & $-0.025490872250634$ & & \\ 
	\botrule
	\end{tabular}
  \label{tab:FITP11Bdm015}
	}
	\end{center}
\end{table}

\begin{table}[H]
	\begin{center}
	\tbl{The numerical values of $p_{1,1}$ with $b_j=c_j=1$ and $m_0=1.0$.}
	{
	\begin{tabular}{lccc} \toprule
	$N_5$ & $8$ & $16$ & $32$ \\ \colrule
	$L/a$& \multicolumn{3}{c}{$p_{1,1}$} \\ \colrule
   $4$ & $-0.000996246152616$ & $-0.000931659891244$ & $-0.000931919758575$ \\ 
   $6$ & $-0.004603150636507$ & $-0.004379720711230$ & $-0.004377564445862$ \\ 
   $8$ & $-0.007570846735406$ & $-0.007339903403487$ & $-0.007334795622206$ \\ 
  $10$ & $-0.009663145567390$ & $-0.009487128771071$ & $-0.009482501467147$ \\ 
  $12$ & $-0.011187581989197$ & $-0.011061057467551$ & $-0.011058231188890$ \\ 
  $14$ & $-0.012384896118215$ & $-0.012290534341752$ & $-0.012289409163346$ \\ 
  $16$ & $-0.013389461252875$ & $-0.013310783001584$ & $-0.013310976363075$ \\ 
  $18$ & $-0.014270981623373$ & $-0.014195328866723$ & $-0.014196493673909$ \\ 
  $20$ & $-0.015065582916720$ & $-0.014984779892570$ & $-0.014986622226234$ \\ 
  $22$ & $-0.015792971302519$ & $-0.015702711408268$ & $-0.015704971170587$ \\ 
  $24$ & $-0.016465000994347$ & $-0.016363707733056$ & $-0.016366158318439$ \\ 
  $26$ & $-0.017089703923897$ & $-0.016977430784739$ & $-0.016979889094112$ \\ 
  $28$ & $-0.017673114155514$ & $-0.017550726401246$ & $-0.017553059566649$ \\ 
  $30$ & $-0.018220085378664$ & $-0.018088745432749$ & $-0.018090869408541$ \\ 
  $32$ & $-0.018734670587965$ & $-0.018595557661083$ & $-0.018597429341229$ \\ 
  $34$ & $-0.019220317416919$ & $-0.019074498753218$ & $-0.019076105281324$ \\ 
  $36$ & $-0.019679985859805$ & $-0.019528374164715$ & $-0.019529722477420$ \\ 
  $38$ & $-0.020116231188214$ & $-0.019959585129973$ & $-0.019960693392692$ \\ 
  $40$ & $-0.020531268659271$ & $-0.020370211346101$ & $-0.020371102773105$ \\ 
  $42$ & $-0.020927026511223$ & $-0.020762068840545$ & $-0.020762767700268$ \\ 
  $44$ & $-0.021305190093525$ & $-0.021136752957334$ & $-0.021137282277683$ \\ 
  $46$ & $-0.021667238733567$ & $-0.021495671861383$ & $-0.021496052316441$ \\ 
  $48$ & $-0.022014476511803$ & $-0.021840073569361$ & $-0.021840323109700$ \\ 
	\botrule
	\end{tabular}
  \label{tab:FITP11Bd}
	}
	\end{center}
\end{table}

\section{Equivalence of the fermionic contribution $p_{1,1}$ between the five-dimensional operator 
and the effective four-dimensional operator}
\label{appendixB}

In this appendix, we prove the equivalence of $p_{1,1}$
constructed with the MDWF and with the effective four-dimensional operators.

With the MDWF operator in the five-dimensional lattice, 
the fermionic contribution $p_{1,1}$ is defined by
\begin{align}
p_{1,1} &= \frac{1}{k} \frac{\partial }{\partial \eta}
    \Bigl[ \mbox{Tr} \ln\left[Z^{-1} D_{PV}^{-1}D_{\MDWF} \right] \Bigr]
\notag\\
&= \frac{1}{k} 
 \mbox{Tr}\left[
 \dfrac{\partial D_{\MDWF}}{\partial \eta}(D_{\MDWF})^{-1} 
-\dfrac{\partial D_{\PV}}{\partial \eta}(D_{\PV})^{-1} 
\right].
\label{eq:SSF_5D}
\end{align}
The trace is taken on the five-dimensional lattice sites, color, and spinor indices.
Eq.~(\ref{eq:SSF}) is the momentum space representation of Eq.~(\ref{eq:SSF_5D}).
In this appendix, we omit the superscript $N_5$
and substitute $\eta=0$ after the differentiation by $\eta$ for simplicity.

The one-loop contribution to the effective action 
induced from the action Eq.~(\ref{eq:truncatedoverlap})
with the effective four-dimensional operator is 
\begin{align}
p^{\mathrm{eff}}_{1,1} &= \frac{1}{k} \frac{\partial }{\partial \eta}
    \Bigl[ \mbox{tr} \ln\left[D_q \right] \Bigr],
\label{eq:SSF_DEF}
\end{align}
where $D_q$ is defined in Eq.~(\ref{eq:renormalizedDq}) 
and the trace is taken on the four-dimensional lattice sites, color, and spinor indices.
The effective four-dimensional operator is renormalized by $Z$ and $am_{\mathrm{res}}$ according to 
Eqs.~(\ref{eq:renormalizedDq}), (\ref{eq:Normalization}), and (\ref{eq:ResMass}).
We show $p_{1,1}=p^{\mathrm{eff}}_{1,1}$ in the following.

Substituting Eqs.~(\ref{eq:EffOP}) and (\ref{eq:renormalizedDq}) into 
 Eq.~(\ref{eq:SSF_DEF}), we have
\begin{align}
p^{\mathrm{eff}}_{1,1} 
 &= \frac{1}{k} \mbox{tr} 
  \left[ 
\epsilon^T P^T
(D_{\MDWF})^{-1}D_{\PV}
P \epsilon  
\right.
\notag\\
&\quad\qquad\times
\left.
\epsilon^T P^T
      \left\{ -(D_{\PV})^{-1}
        \frac{\partial D_{\PV}}{\partial \eta} D_{\PV}^{-1} D_{\MDWF}
        + (D_{\PV})^{-1} \frac{\partial D_{\MDWF}}{\partial \eta}
      \right\} P \epsilon  
  \right]. 
    \label{eq:SSF4D}
\end{align}

We introduce two matrices $A$ and $B$ defined by
\begin{align}
  A &= P^T (D_{\MDWF})^{-1} D_{\PV} P 
\label{eq:efov5d}, 
\\
  B &= P^T \left\{ (D_{\PV})^{-1} \frac{\partial D_{\MDWF}}{\partial \eta} 
                  -(D_{\PV})^{-1} \frac{\partial D_{\PV}  }{\partial \eta} 
                   (D_{\PV})^{-1} D_{\MDWF}\right\} P.
\label{eq:derv5d}
\end{align}
to simplify $p_{1,1}$ and $p^{\mathrm{eff}}_{1,1}$ as
\begin{align}
                 p_{1,1} &= \frac{1}{k} \mathrm{Tr}\left[AB\right], 
\label{eq:P115DAB}
\\
  p^{\mathrm{eff}}_{1,1} &= \frac{1}{k} \mathrm{tr} 
    \left[ (\epsilon^T A \epsilon) \; (\epsilon^T B \epsilon) \right],
\label{eq:P114DAB}
\end{align}
where $PP^{T}=1$ is used.

After some matrix algebra in the five-dimensional notation, we obtain
\begin{align}
A & =
\arraycolsep5pt
 \left(
   \begin{array}{@{\,}c|cccc@{\,}}
     1 - (1-m_f)\alpha & & 0 & & \\  \hline
     & & & & \\
     \overrightarrow{z} & \multicolumn{4}{c}{\raisebox{-10pt}[0pt][0pt]{\Huge $\bm{I}$}} \\
     & & & & \\
   \end{array}
 \right),
\label{eq:mat_efov}
\\
B &= 
\arraycolsep5pt
  \left(
    \begin{array}{@{\,}c|cccc@{\,}}
      \gamma & & 0 & & \\ \hline
             & & & & \\
\overrightarrow{g} & \multicolumn{4}{c}{\raisebox{-10pt}[0pt][0pt]{\Huge $\bm{0}$}} \\
             & & & & \\
    \end{array}
  \right),
  \label{eq:mat_derv}
\end{align}
where $\alpha$ and $\gamma$ are four-dimensional operators and $\vec{z}$ and $\vec{g}$ contain
$N_5-1$ four-dimensional operators. Their explicit forms are not required in the proof.

Substituting 
Eqs.~(\ref{eq:mat_efov}) and (\ref{eq:mat_derv}) into Eq.~(\ref{eq:P115DAB}), we obtain
\begin{align}
  p_{1,1} &= \frac{1}{k} \mathrm{Tr}\Biggl[
  \arraycolsep5pt
  \left(
    \begin{array}{@{\,}c|cccc@{\,}}
      1 - (1-m_f)\alpha & & 0 & & \\\hline
      & & & & \\
      \overrightarrow{z} & \multicolumn{4}{c}{\raisebox{-10pt}[0pt][0pt]{\Huge $\bm{I}$}} \\
      & & & & \\
    \end{array}
  \right)
  \arraycolsep5pt
  \left(
    \begin{array}{@{\,}c|cccc@{\,}}
      \gamma & & 0 & & \\\hline
      & & & & \\
      \overrightarrow{g} & \multicolumn{4}{c}{\raisebox{-10pt}[0pt][0pt]{\Huge $\bm{0}$}} \\
      & & & & \\
    \end{array}
  \right) \Biggr]
 \notag \\
   &= \frac{1}{k} \mathrm{tr} [ \{ 1 - (1-m_f)\alpha \}\gamma ],
  \label{eq:P115D}
\end{align}
where the trace in the last line is now taken only on the four-dimensional lattice, color, and spinor indices.

Similarly substituting 
Eqs.~(\ref{eq:mat_efov}) and (\ref{eq:mat_derv})  into Eq.~(\ref{eq:P114DAB}), we obtain
\begin{align}
p^{\mathrm{eff}}_{1,1}
  &= \frac{1}{k} \mathrm{Tr} 
\Biggl[
  \begin{pmatrix}
  1 & 0 &\cdots & 0 
  \end{pmatrix}
  \arraycolsep5pt
  \left(
  \begin{array}{@{\,}c|cccc@{\,}}
    1 - (1-m_f)\alpha & & 0 & & \\
    \hline
    & & & & \\
    \overrightarrow{z} & \multicolumn{4}{c}{\raisebox{-10pt}[0pt][0pt]{\Huge $\bm{I}$}} \\
    & & & & \\
  \end{array}
  \right)
  \begin{pmatrix}
    1 \\  0 \\ \vdots \\  0 
  \end{pmatrix}
\notag\\
& \qquad \qquad \qquad \qquad \qquad \times
  \begin{pmatrix}
    1 & 0 &\cdots & 0 
  \end{pmatrix}
  \arraycolsep5pt
  \left(
  \begin{array}{@{\,}c|cccc@{\,}}
    \gamma & &0 & & \\\hline
           & & & & \\
    \overrightarrow{g}&\multicolumn{4}{c}{\raisebox{-10pt}[0pt][0pt]{\Huge $\bm{0}$}} \\
           & & & & \\
    \end{array}
  \right)
  \begin{pmatrix}
    1 \\ 0 \\ \vdots \\ 0 
  \end{pmatrix} 
\Biggr]
  \notag \\
  & = \frac{1}{k} \mathrm{tr} [ \{ 1 - (1-m_f)\alpha \}\gamma ].
  \label{eq:P114D}
\end{align}
Thus, we have proved $p_{1,1}=p^{\mathrm{eff}}_{1,1}$.

We also note that 
\begin{align}
    (D_{q})^{-1} = \epsilon^{T} P^{T} (D_{\MDWF})^{-1}D_{\PV} P \epsilon,
\end{align}
holds even with the SF boundary term.


\begin{thebibliography}{10}

\bibitem{Hasenfratz:1998jp} 
  P.~Hasenfratz,
  Nucl.\ Phys.\ B {\bf 525}, 401 (1998)
  [hep-lat/9802007].
\bibitem{Hasenfratz:1998ri} 
  P.~Hasenfratz, V.~Laliena and F.~Niedermayer,
  Phys.\ Lett.\ B {\bf 427}, 125 (1998)
  [hep-lat/9801021].

\bibitem{Kaplan:1992bt} 
  D.~B.~Kaplan,
  Phys.\ Lett.\ B {\bf 288}, 342 (1992)
  [hep-lat/9206013].

\bibitem{Narayanan:1992wx} 
  R.~Narayanan and H.~Neuberger,
  Phys.\ Lett.\ B {\bf 302}, 62 (1993)
  [hep-lat/9212019].
\bibitem{Narayanan:1993ss} 
  R.~Narayanan and H.~Neuberger,
  Phys.\ Rev.\ Lett.\  {\bf 71}, 3251 (1993)
  [hep-lat/9308011].
\bibitem{Narayanan:1993sk} 
  R.~Narayanan and H.~Neuberger,
  Nucl.\ Phys.\ B {\bf 412}, 574 (1994)
  [hep-lat/9307006].

\bibitem{Ginsparg:1981bj} 
  P.~H.~Ginsparg and K.~G.~Wilson,
  Phys.\ Rev.\ D {\bf 25}, 2649 (1982).

\bibitem{Luscher:1998pqa} 
  M.~L\"{u}scher,
  Phys.\ Lett.\ B {\bf 428}, 342 (1998)
  [hep-lat/9802011].

\bibitem{Creutz:2000bs} 
  M.~Creutz,
  Rev.\ Mod.\ Phys.\  {\bf 73}, 119 (2001)
  [hep-lat/0007032].
\bibitem{Neuberger:2001nb} 
  H.~Neuberger,
  Ann.\ Rev.\ Nucl.\ Part.\ Sci.\  {\bf 51}, 23 (2001)
  [hep-lat/0101006].
\bibitem{Chandrasekharan:2004cn} 
  S.~Chandrasekharan and U.~J.~Wiese,
  Prog.\ Part.\ Nucl.\ Phys.\  {\bf 53}, 373 (2004)
  [hep-lat/0405024].


\bibitem{RECENTQCDRESULTS}
  S.~Aoki {\it et al.},
  {\it Eur.\ Phys.\ J.\ C} {\bf 74}, 2890 (2014),
  arXiv:1310.8555 [hep-lat].

\bibitem{Martinelli:1993dq} 
  G.~Martinelli, S.~Petrarca, C.~T.~Sachrajda and A.~Vladikas,
  Phys.\ Lett.\ B {\bf 311}, 241 (1993)
  [Erratum-ibid.\ B {\bf 317}, 660 (1993)].
\bibitem{Martinelli:1994ty} 
  G.~Martinelli, C.~Pittori, C.~T.~Sachrajda, M.~Testa and A.~Vladikas,
  Nucl.\ Phys.\ B {\bf 445}, 81 (1995)
  [hep-lat/9411010].

\bibitem{Luscher:1992an} 
  M.~L\"{u}scher, R.~Narayanan, P.~Weisz and U.~Wolff,
  Nucl.\ Phys.\ B {\bf 384}, 168 (1992)
  [hep-lat/9207009].
\bibitem{Sint:1993un} 
  S.~Sint,
  Nucl.\ Phys.\ B {\bf 421}, 135 (1994)
  [hep-lat/9312079].
\bibitem{Luscher:1993gh} 
  M.~L\"{u}scher, R.~Sommer, P.~Weisz and U.~Wolff,
  Nucl.\ Phys.\ B {\bf 413}, 481 (1994)
  [hep-lat/9309005].
\bibitem{Sint:1995rb} 
  S.~Sint,
  Nucl.\ Phys.\ B {\bf 451}, 416 (1995)
  [hep-lat/9504005].

\bibitem{Luscher:2010iy} 
  M.~L\"{u}scher,
  JHEP {\bf 1008}, 071 (2010)
  [Erratum-ibid.\  {\bf 1403}, 092 (2014)]
  [arXiv:1006.4518 [hep-lat]].
\bibitem{Luscher:2011bx} 
  M.~L\"{u}scher and P.~Weisz,
  JHEP {\bf 1102}, 051 (2011)
  [arXiv:1101.0963 [hep-th]].
\bibitem{Fodor:2012td} 
  Z.~Fodor, K.~Holland, J.~Kuti, D.~Nogradi and C.~H.~Wong,
  JHEP {\bf 1211}, 007 (2012)
  [arXiv:1208.1051 [hep-lat]].
\bibitem{Fritzsch:2013je} 
  P.~Fritzsch and A.~Ramos,
  JHEP {\bf 1310}, 008 (2013)
  [arXiv:1301.4388 [hep-lat]].
\bibitem{Suzuki:2013gza} 
  H.~Suzuki,
  PTEP {\bf 2013}, no. 8, 083B03 (2013)
  [arXiv:1304.0533 [hep-lat]].
\bibitem{Luscher:2013cpa} 
  M.~L\"{u}scher,
  JHEP {\bf 1304}, 123 (2013)
  [arXiv:1302.5246 [hep-lat]].

\bibitem{Luscher:1996vw} 
  M.~L\"{u}scher and P.~Weisz,
  Nucl.\ Phys.\ B {\bf 479}, 429 (1996)
  [hep-lat/9606016].
\bibitem{Luscher:1996ug} 
  M.~L\"{u}scher, S.~Sint, R.~Sommer, P.~Weisz and U.~Wolff,
  Nucl.\ Phys.\ B {\bf 491}, 323 (1997)
  [hep-lat/9609035].
\bibitem{Luscher:1996jn} 
  M.~L\"{u}scher, S.~Sint, R.~Sommer and H.~Wittig,
  Nucl.\ Phys.\ B {\bf 491}, 344 (1997)
  [hep-lat/9611015].
\bibitem{Jansen:1998mx} 
  K.~Jansen {\it et al.}  [ALPHA Collaboration],
  Nucl.\ Phys.\ B {\bf 530}, 185 (1998)
  [Erratum-ibid.\ B {\bf 643}, 517 (2002)]
  [hep-lat/9803017].
\bibitem{Capitani:1998mq} 
  S.~Capitani {\it et al.}  [ALPHA Collaboration],
  Nucl.\ Phys.\ B {\bf 544}, 669 (1999)
  [hep-lat/9810063].
\bibitem{Sint:1998iq} 
  S.~Sint {\it et al.}  [ALPHA Collaboration],
  Nucl.\ Phys.\ B {\bf 545}, 529 (1999)
  [hep-lat/9808013].
\bibitem{Bode:2001jv} 
  A.~Bode {\it et al.}  [ALPHA Collaboration],
  Phys.\ Lett.\ B {\bf 515}, 49 (2001)
  [hep-lat/0105003].
\bibitem{DellaMorte:2004bc} 
  M.~Della Morte {\it et al.}  [ALPHA Collaboration],
  Nucl.\ Phys.\ B {\bf 713}, 378 (2005)
  [hep-lat/0411025].
\bibitem{Guagnelli:2005zc} 
  M.~Guagnelli {\it et al.}  [ALPHA Collaboration],
  JHEP {\bf 0603}, 088 (2006)
  [hep-lat/0505002].
\bibitem{DellaMorte:2005kg} 
  M.~Della Morte {\it et al.}  [ALPHA Collaboration],
  Nucl.\ Phys.\ B {\bf 729}, 117 (2005)
  [hep-lat/0507035].
\bibitem{Dimopoulos:2007ht} 
  P.~Dimopoulos {\it et al.}  [ALPHA Collaboration],
  JHEP {\bf 0805}, 065 (2008)
  [arXiv:0712.2429 [hep-lat]].


\bibitem{Taniguchi:2006qw} 
  Y.~Taniguchi,
  JHEP {\bf 0610}, 027 (2006)
  [hep-lat/0604002].

\bibitem{Taniguchi:2004gf} 
  Y.~Taniguchi,
  JHEP {\bf 0512}, 037 (2005)
  [hep-lat/0412024].

\bibitem{Luscher:2006df} 
  M.~L\"{u}scher,
  JHEP {\bf 0605}, 042 (2006)
  [hep-lat/0603029].

\bibitem{Takeda:2007ga} 
  S.~Takeda,
  Nucl.\ Phys.\ B {\bf 796}, 402 (2008)
  [arXiv:0712.1469 [hep-lat]].

\bibitem{Takeda:2010ai} 
  S.~Takeda,
  Phys.\ Rev.\ D {\bf 87}, no. 11, 114506 (2013)
  [arXiv:1010.3504 [hep-lat]].

\bibitem{Sint:2007zz} 
  S.~Sint,
  PoS LAT {\bf 2007}, 253 (2007).

\bibitem{Brower:2004xi} 
  R.~C.~Brower, H.~Neff and K.~Orginos,
  Nucl.\ Phys.\ Proc.\ Suppl.\  {\bf 140}, 686 (2005)
  [hep-lat/0409118].
\bibitem{Brower:2012vk} 
  R.~C.~Brower, H.~Neff and K.~Orginos,
  arXiv:1206.5214 [hep-lat].

\bibitem{Hashimoto:2014gta} 
  S.~Hashimoto, S.~Aoki, G.~Cossu, H.~Fukaya, T.~Kaneko, J.~Noaki and P.~A.~Boyle,
  PoS LATTICE {\bf 2013}, 431 (2014).


\bibitem{Murakami:2014qfa} 
  Y.~Murakami and K.~I.~Ishikawa,
  PoS LATTICE {\bf 2014}, 331 (2014)
  [arXiv:1410.8335 [hep-lat]].

\bibitem{Murakami:2017wcz} 
  Y.~Murakami and K.~I.~Ishikawa,
  PoS LATTICE {\bf 2015}, 308 (2016)
  [arXiv:1701.07139 [hep-lat]].

 \bibitem{Kikukawa:1999sy} 
  Y.~Kikukawa and T.~Noguchi,
  hep-lat/9902022.

\bibitem{Kikukawa:1999dk}
  Y.~Kikukawa,
  Nucl.\ Phys.\ B {\bf 584}, 511 (2000)
  [hep-lat/9912056].

\bibitem{Capitani:2007ez} 
  S.~Capitani,
  PoS LAT {\bf 2007}, 066 (2007)
  [arXiv:0708.3281 [hep-lat]].

  \bibitem{Borici:1999zw}
  A.~Bori\c{c}i,
  Nucl.\ Phys.\ Proc.\ Suppl.\  {\bf 83} (2000) 771 [hep-lat/9909057].

  \bibitem{Borici:1999da} 
  A.~Bori\c{c}i,
  in Lattice Fermions and Structure of the Vacuum, V. Mitrjushkin {\it et al.} (edts.), Springer 2000, [hep-lat/9912040].

  \bibitem{Borici:2004pn} 
  A.~Bori\c{c}i,
  in QCD and numerical analysis III, A.~Bori\c{c}i {\it et al.} (edts.), Springer 2005, [hep-lat/0402035].

 \bibitem{Chiu:2002ir} 
  T.~W.~Chiu,
  Phys.\ Rev.\ Lett.\  {\bf 90}, 071601 (2003)
  [hep-lat/0209153].

 \bibitem{Furman:1994ky} 
  V.~Furman and Y.~Shamir,
  Nucl.\ Phys.\ B {\bf 439}, 54 (1995)
  [hep-lat/9405004].

 \bibitem{Chen:2014hyy} 
  Y.~C.~Chen and T.~W.~Chiu,
  Phys.\ Lett.\ B {\bf 738}, 55 (2014)
  [arXiv:1403.1683 [hep-lat]].
 \bibitem{Ogawa:2009ex} 
  K.~Ogawa {\it et al.}  [TWQCD Collaboration],
  PoS LAT {\bf 2009}, 033 (2009)
  [arXiv:0911.5532 [hep-lat]].

\bibitem{Boyle:2015vda}
  P.~A.~Boyle [UKQCD Collaboration],
  PoS LATTICE {\bf 2014} (2015) 087.

\bibitem{Chiu:2015sea}
  T.~W.~Chiu,
  Phys.\ Lett.\ B {\bf 744} (2015) 95
  [arXiv:1503.01750 [hep-lat]].

 \bibitem{Chiu:2002eh} 
  T.~W.~Chiu, T.~H.~Hsieh, C.~H.~Huang and T.~R.~Huang,
  Phys.\ Rev.\ D {\bf 66}, 114502 (2002)
  [hep-lat/0206007].

\bibitem{textbook:rothe}
  H.~J.~Rothe,
  {\it World Sci.\ Lect.\ Notes Phys.\ } {\bf 43}, 1 (1992)
   [{\it World Sci.\ Lect.\ Notes Phys.\ } {\bf 59}, 1 (1997); {\it ibid.} {\bf 74}, 1 (2005);  {\bf 82}, 1 (2012)].

\bibitem{textbook:itzyksonzuber}
  C.~Itzykson and J.~B.~Zuber, 
  ``Quantum Field Theory,'' McGraw-Hill Book Co., New York, 1988.

\bibitem{Sint:1995ch}
  S.~Sint and R.~Sommer,
  Nucl.\ Phys.\ B {\bf 465} (1996) 71
  [hep-lat/9508012].

\bibitem{Sint:2010eh}
  S.~Sint,
  Nucl.\ Phys.\ B {\bf 847} (2011) 491
  [arXiv:1008.4857 [hep-lat]].


\bibitem{Luscher:1996sc}
  M.~Luscher, S.~Sint, R.~Sommer and P.~Weisz,
  Nucl.\ Phys.\ B {\bf 478} (1996) 365
  [hep-lat/9605038].

\bibitem{Aoki:2003uf}
  S.~Aoki and Y.~Kuramashi,
  Phys.\ Rev.\ D {\bf 68} (2003) 034507
  [hep-lat/0306008].

\bibitem{Alexandrou:1999wr}
  C.~Alexandrou, H.~Panagopoulos and E.~Vicari,
  Nucl.\ Phys.\ B {\bf 571} (2000) 257
  [hep-lat/9909158].

\bibitem{Luscher:1991wu} 
  M.~L\"{u}scher, P.~Weisz and U.~Wolff,
  Nucl.\ Phys.\ B {\bf 359}, 221 (1991).


\end{thebibliography}
\end{document}